\documentclass[aip,amsmath,amssymb,preprint]{revtex4-1}
\usepackage[a-1b]{pdfx}
\usepackage[utf8]{inputenc}
\usepackage{amsfonts}
\usepackage{amsmath}
\usepackage{amssymb}
\usepackage{graphicx}
\usepackage{multirow}
\usepackage{listings}
\usepackage{ragged2e}
\usepackage{xfp}
\usepackage{xcolor}         %
\usepackage{color}
\usepackage{bm}
\usepackage{longtable}

\PassOptionsToPackage{hyphens}{url}\usepackage{hyperref}

\newcounter{bla}

\usepackage{epstopdf}
\AppendGraphicsExtensions{.tif} 

\usepackage[scientific-notation=false]{siunitx}

\newcommand{\float}[1]{\num[round-integer-to-decimal,round-mode=figures,round-precision=3]{\fpeval{#1 * 1000}}}
\newcommand{\floatm}[1]{\num[round-integer-to-decimal,round-mode=figures,round-precision=3]{\fpeval{#1 * 1000000}}}

\usepackage{etoolbox}
\makeatletter
\def\@email#1#2{%
 \endgroup
 \patchcmd{\titleblock@produce}
  {\frontmatter@RRAPformat}
  {\frontmatter@RRAPformat{\produce@RRAP{*#1\href{mailto:#2}{#2}}}\frontmatter@RRAPformat}
  {}{}
}
\usepackage{CJKutf8}
\CJKnospace
\makeatother

\begin{document}
\begin{CJK*}{UTF8}{gbsn}
\title{DeePMD-kit v2: A software package for Deep Potential models}

\author{Jinzhe Zeng}%
\affiliation{Laboratory for Biomolecular Simulation Research, Institute for Quantitative Biomedicine and Department of Chemistry and Chemical Biology, Rutgers University, Piscataway, New Jersey 08854, United States}

\author{Duo Zhang}%
\affiliation{AI for Science Institute, Beijing 100080, P.R.~China}
\affiliation{DP Technology, Beijing 100080, P.R.~China}
\affiliation{Academy for Advanced Interdisciplinary Studies, Peking University, Beijing 100871, P.R.~China}

\author{Denghui Lu}%
\affiliation{HEDPS, CAPT, College of Engineering, Peking University, Beijing 100871, P.R.~China}

\author{Pinghui Mo}%
\affiliation{College of Electrical and Information Engineering, Hunan University, Changsha, P.R.~China}

\author{Zeyu Li}%
\affiliation{Yuanpei College, Peking University, Beijing 100871, P.R.~China}

\author{Yixiao Chen}%
\affiliation{Program in Applied and Computational Mathematics, Princeton University, Princeton, New Jersey 08540, United States}

\author{Marián Rynik}
\affiliation{Department of Experimental Physics, Comenius University, Mlynská Dolina F2, 842 48 Bratislava, Slovakia}

\author{Li'ang Huang}%
\affiliation{Center for Quantum Information, Institute for Interdisciplinary Information Sciences, Tsinghua University, Beijing 100084, P.R.~China}

\author{Ziyao Li}%
\affiliation{Center for Data Science, Peking University, Beijing 100871, P.R.~China}
\affiliation{DP Technology, Beijing 100080, P.R.~China}

\author{Shaochen Shi}
\affiliation{ByteDance Research, Zhonghang Plaza, No. 43, North 3rd Ring West Road, Haidian District, Beijing, P.R.~China}

\author{Yingze Wang }%
\affiliation{College of Chemistry and Molecular Engineering, Peking University, Beijing 100871, P.R.~China}
\affiliation{DP Technology, Beijing 100080, P.R.~China}

\author{Haotian Ye}%
\affiliation{Yuanpei College, Peking University, Beijing 100871, P.R.~China}

\author{Ping Tuo}%
\affiliation{AI for Science Institute, Beijing 100080, P.R.~China}

\author{Jiabin Yang }%
\affiliation{Baidu Inc., Beijing, P.R.~China}

\author{Ye Ding }%
\affiliation{Key Laboratory of Structural Biology of Zhejiang Province, School of Life Sciences, Westlake University, Hangzhou, Zhejiang, P.R.~China}
\affiliation{Westlake AI Therapeutics Lab, Westlake Laboratory of Life Sciences and Biomedicine, Hangzhou, Zhejiang, P.R.~China}

\author{Yifan Li}%
\affiliation{Department of Chemistry, Princeton University, Princeton, New Jersey 08544, United States}

\author{Davide Tisi}
\affiliation{SISSA, Scuola Internazionale Superiore di Studi Avanzati, 34136,  Trieste, Italy}
\affiliation{Laboratory of Computational Science and Modeling, Institute of  Materials, École Polytechnique Fédérale de Lausanne, 1015 Lausanne, Switzerland}

\author{Qiyu Zeng}%
\affiliation{Department of Physics, National University of Defense Technology, Changsha, Hunan 410073, P.R.~China}

\author{Han Bao}
\affiliation{State Key Lab of Processors, Institute of Computing Technology, Chinese Academy of Sciences, Beijing, P.R.~China}
\affiliation{University of Chinese Academy of Sciences, Beijing, P.R.~China}

\author{Yu Xia }%
\affiliation{ByteDance Research, Zhonghang Plaza, No. 43, North 3rd Ring West Road, Haidian District, Beijing, P.R.~China}

\author{Jiameng Huang }%
\affiliation{DP Technology, Beijing 100080, P.R.~China}
\affiliation{School of Electronics Engineering and Computer Science, Peking University, Beijing 100871, P.R.~China}

\author{Koki Muraoka}
\affiliation{Department of Chemical System Engineering, The University of Tokyo, 7-3-1 Hongo, Bunkyo-ku, Tokyo 113-8656, Japan}

\author{Yibo Wang}%
\affiliation{DP Technology, Beijing 100080, P.R.~China}

\author{Junhan Chang }%
\affiliation{DP Technology, Beijing 100080, P.R.~China}
\affiliation{College of Chemistry and Molecular Engineering, Peking University, Beijing 100871, P.R.~China}

\author{Fengbo Yuan }%
\affiliation{DP Technology, Beijing 100080, P.R.~China}

\author{Sigbjørn Løland Bore}
\affiliation{Hylleraas Centre for Quantum Molecular Sciences and Department of Chemistry, University of Oslo, PO Box 1033 Blindern, 0315 Oslo, Norway}

\author{Chun Cai }%
\affiliation{AI for Science Institute, Beijing 100080, P.R.~China}
\affiliation{DP Technology, Beijing 100080, P.R.~China}

\author{Yinnian Lin }%
\affiliation{Wangxuan Institute of Computer Technology, Peking University, Beijing 100871, P.R.~China}

\author{Bo Wang }%
\affiliation{Shanghai Engineering Research Center of Molecular Therapeutics \& New Drug Development, Shanghai Key Laboratory of Green Chemistry \& Chemical Process, School of Chemistry and Molecular Engineering, East China Normal University, Shanghai 200062, P.R.~China}

\author{Jiayan Xu }%
\affiliation{School of Chemistry and Chemical Engineering, Queen’s University Belfast, Belfast BT9 5AG, U.K.}

\author{Jia-Xin Zhu}%
\affiliation{State Key Laboratory of Physical Chemistry of Solid Surfaces, iChEM, College of Chemistry and Chemical Engineering, Xiamen University, Xiamen 361005, P.R.~China}

\author{Chenxing Luo }%
\affiliation{Department of Applied Physics and Applied Mathematics, Columbia University, New York, NY 10027, United States}

\author{Yuzhi Zhang }%
\affiliation{DP Technology, Beijing 100080, P.R.~China}

\author{Rhys E. A. Goodall}
\affiliation{Independent Researcher, London, UK}

\author{Wenshuo Liang}
\affiliation{DP Technology, Beijing 100080, P.R.~China}

\author{Anurag Kumar Singh}
\affiliation{Department of Data Science, Indian Institute of Technology Palakkad, Kerala, India}

\author{Sikai Yao }%
\affiliation{DP Technology, Beijing 100080, P.R.~China}

\author{Jingchao Zhang }%
\affiliation{NVIDIA AI Technology Center (NVAITC), Santa Clara, CA 95051, United States}

\author{Renata Wentzcovitch}
\affiliation{Department of Applied Physics and Applied Mathematics, Columbia University, New York, NY 10027, United States}
\affiliation{Department of Earth and Environmental Sciences, Columbia University, New York, NY 10027, United States}

\author{Jiequn Han}
\affiliation{Center for Computational Mathematics, Flatiron Institute, New York, NY 10010, United States}

\author{Jie Liu}
\affiliation{College of Electrical and Information Engineering, Hunan University, Changsha, P.R.~China}

\author{Weile Jia }%
\affiliation{State Key Lab of Processors, Institute of Computing Technology, Chinese Academy of Sciences, Beijing, P.R.~China}
\affiliation{University of Chinese Academy of Sciences, Beijing, P.R.~China}

\author{Darrin M. York}
\affiliation{Laboratory for Biomolecular Simulation Research, Institute for Quantitative Biomedicine and Department of Chemistry and Chemical Biology, Rutgers University, Piscataway, New Jersey 08854, United States}

\author{Weinan E}%
\affiliation{Center for Machine Learning Research and School of Mathematical Sciences, Peking University, Beijing 100871, People’s Republic of China}
\affiliation{AI for Science Institute, Beijing 100080, P.R.~China}

\author{Roberto Car}
\affiliation{Department of Chemistry, Princeton University, Princeton, New Jersey 08544, United States}

\author{Linfeng Zhang }%
\email{linfeng.zhang.zlf@gmail.com}
\affiliation{DP Technology, Beijing 100080, P.R.~China}
\affiliation{AI for Science Institute, Beijing 100080, P.R.~China}

\author{Han Wang }%
\email{wang\_han@iapcm.ac.cn}
\affiliation{Laboratory of Computational Physics, Institute of Applied Physics and Computational Mathematics, Fenghao East Road 2, Beijing 100094, P.R.~China}
\affiliation{HEDPS, CAPT, College of Engineering, Peking University, Beijing 100871, P.R.~China}

\begin{abstract}

DeePMD-kit is a powerful open-source software package that facilitates molecular dynamics simulations using machine learning potentials (MLP) known as Deep Potential (DP) models.
This package, which was released in 2017, has been widely used in the fields of physics, chemistry, biology, and material science for studying atomistic systems.
The current version of DeePMD-kit offers numerous advanced features such as DeepPot-SE, attention-based and hybrid descriptors, the ability to fit tensile properties, type embedding, model deviation, Deep Potential - Range Correction (DPRc), Deep Potential Long Range (DPLR), GPU support for customized operators, model compression, non-von Neumann molecular dynamics (NVNMD), and improved usability, including documentation, compiled binary packages, graphical user interfaces (GUI), and application programming interfaces (API).
This article presents an overview of the current major version of the DeePMD-kit package, highlighting its features and technical details.
Additionally, the article benchmarks the accuracy and efficiency of different models and discusses ongoing developments.

\end{abstract}

\maketitle
\end{CJK*}

\section{Introduction}

In recent years, the increasing popularity of machine learning potentials (MLP) has revolutionized molecular dynamics (MD) simulations across various fields.
\cite{Behler_PhysRevLett_2007_v98_p146401,Bartok_PhysRevLett_2010_v104_p136403,Behler_JChemPhys_2011_v134_p074106,Gastegger_JChemPhys_2018_v148_p241709,Chmiela_SciAdv_2017_v3_p1603015,Schutt_NatCommun_2017_v8_p13890,Schutt_JChemPhys_2018_v148_p241722,Chen_JChemTheoryComput_2018_v14_p3933,Zhang_PhysRevLett_2018_v120_p143001,Zhang_BookChap_NIPS_2018_v31_p4436,Zhang_JPhysChemLett_2019_v10_p4962,Smith_ChemSci_2017_v8_p3192,Unke_JChemTheoryComput_2019_v15_p3678,GLick_JChemPhys_2020_v153_p044112,Zubatiuk_AccChemRes_2021_v54_p1575,Khajehpasha_PhysRevB_2022_v105_p144106,Pan_JChemTheoryComput_2021_v17_p5745,takamoto_towards_2022}
Numerous software packages have been developed to support the use of MLPs.
\cite{Wang_ComputPhysCommun_2018_v228_p178,Schutt_JChemTheoryComput_2019_v15_p448,Chmiela_ComputPhysCommun_2019_v240_p38,Unke_JChemTheoryComput_2019_v15_p3678,Lee_ComputPhysCommun_2019_v242_p95,Gao_JChemInfModel_2020_v60_p3408,Chmiela_ComputerPhysicsCommunications_2019_v240_p38,Dral_TopCurrChem(Cham)_2021_v379_p27,Singraber_JChemTheoryComput_2019_v15_p1827,Zhang_JChemPhys_2022_v156_p114801,schutt2022schnetpack,Fan_JChemPhys_2022_v157_p114801,Novikov_MachLearnSciTechnol_2021_v2_p025002,Yanxon_MachLearnSciTechnol_2021_v2_p027001}
One of the main reasons for the widespread adoption of MLPs is their exceptional speed and accuracy, which outperforms traditional molecular mechanics (MM) and \textit{ab initio} quantum mechanics (QM) methods.\cite{Jia_SC20_2020,Guo_PPoPP_2022_p205}
As a result, MLP-powered MD simulations have become ubiquitous in the field and are increasingly recognized as a valuable tool for studying atomistic systems.
\cite{Behler_JChemPhys_2016_v145_p170901,Butler_Nature_2018_v559_p547,Noe_AnnuRevPhysChem_2020_v71_p361,Unke_ChemRev_2021_v121_p10142,Pinheiro_ChemSci_2021_v12_p14396,Manzhos_ChemRev_2021_v121_p10187,Zeng_BookChap_QuantChemML_2022_p279}

DeePMD-kit is an open-source software package that facilitates molecular dynamics (MD) simulations using machine learning potentials (MLPs).
The package was first released in 2017\cite{Wang_ComputPhysCommun_2018_v228_p178} and has since undergone rapid development with contributions from many developers.
DeePMD-kit implements a series of MLP models known as Deep Potential (DP) models,\cite{Zhang_PhysRevLett_2018_v120_p143001,Zhang_BookChap_NIPS_2018_v31_p4436,Wang_NuclFusion_2022_v62_p126013,Zhang_2022_DPA1,Zeng_JChemTheoryComput_2021_v17_p6993,Zhang_JChemPhys_2022_v156_p124107,Liang_BookCharp_MultiscaleModeling_2023_p6-1}
which have been widely adopted in the fields of physics, chemistry, biology, and material science for studying a broad range of atomistic systems.
These systems include
metallic materials\cite{Chen_ACSNano_2021_v15_p12418,Dai_JMatSciTech_2020_v43_p168,Ding_MaterialsScienceinSemiconductorProcessing_2022_v143_p106513,Jiao_AdvSci_2022_v9_pe2105574,Li_ApplPhysLett_2020_v117_p152102,Liu_JPhysCondensMatter_2020_v32_p144002,Niu_NatCommun_2020_v11_p2654,Wu_PhysRevB_2021_v104_p174107,Miyagawa_ComputationalMaterialsScience_2022_v206_p111303,Liang_AdvTheorySimul_2020_v3_p2000180,Pan_ComputationalMaterialsScience_2020_v185_p109955,Dai_JournaloftheEuropeanCeramicSociety_2020_v40_p5029,Wang_FrontChem_2020_v8_p589795,Rodriguez_ACSApplMaterInterfaces_2021_v13_p55367,Wen_npjComputMater_2021_v7_p206,Gupta_EnergyEnvironSci_2021_v14_p6554},
non-metallic inorganic materials\cite{Achar_JPhysChemC_2021_v125_p14874,Bonati_PhysRevLett_2018_v121_p265701,Wang_Carbon_2022_v186_p1,Li_MaterialsTodayPhysics_2020_v12_p100181,Balyakin_PhysRevE_2020_v102_p052125},
water\cite{Ko_MolecularPhysics_2019_v117_p3269,Xu_PhysRevB_2020_v102_p214113,Andreani_JPhysChemLett_2020_v11_p9461,Zhang_PhysRevB_2020_v102_p115155,Gartner_ProcNatlAcadSciUSA_2020_v117_p26040,Tisi_PhysRevB_2021_v104_p224202,Malosso2022_npj,Shi_JPhysChemLett_2021_v12_p10310,Matusalem_ProcNatlAcadSciUSA_2022_v119_pe2203397119,zhai2022shortJCP,bore_paesani_2023}, 
organic systems,\cite{Zhang_BookChap_NIPS_2018_v31_p4436,Zeng_JChemTheoryComput_2023_v19_p1261}
solutions\cite{Zhang_NatCommun_2022_v13_p822,Yang_CatalysisToday_2022_v387_p143,Zeng_JChemTheoryComput_2021_v17_p6993,Giese_JChemTheoryComput_2022_v18_p4304,Liu_PhysChemChemPhys_2023_vNone_pNone},
gas-phase systems\cite{Zeng_NatCommun_2020_v11_p5713,Zeng_EnergyFuels_2021_v35_p762,Chu_JPhysChemLett_2022_v13_p4052,Wang_ChemRxiv_2022},
macromolecular systems,\cite{Wang_JPhysChemB_2020_v124_p3027,Han_BriefBioinform_2021_v22_pNone}
and interfaces\cite{CalegariAndrade_ChemSci_2020_v11_p2335,Galib_Science_2021_v371_p921,Zhuang_JChemPhys_2022_v157_p164701,delaPuente_JAmChemSoc_2022_vNone_pNone,Niblett_JChemPhys_2021_v155_p164101}.
Furthermore, DeePMD-kit is capable of simulating systems containing
almost all periodic table elements\cite{Zhang_2022_DPA1},
operating under a wide range of temperature and pressure,\cite{Zhang_PhysRevLett_2021_v126_p236001}
and can handle
drug-like molecules,\cite{Zeng_JChemTheoryComput_2023_v19_p1261,Zeng_JChemPhys_2023_v158_p124110}
ions,\cite{Zhang_NatCommun_2022_v13_p822,Liu_PhysChemChemPhys_2023_vNone_pNone}
transition states,\cite{Zeng_NatCommun_2020_v11_p5713,Giese_JChemTheoryComput_2022_v18_p4304}
and excited states.\cite{Chen_JPhysChemLett_2018_v9_p6702}
As a result, DeePMD-kit is a powerful and versatile tool that can be used to simulate a wide range of atomistic systems.

Compared to its initial release\cite{Wang_ComputPhysCommun_2018_v228_p178}, DeePMD-kit has evolved significantly, with the current version (v2.2.1) offering an extensive range of features.
These include
DeepPot-SE, attention-based, and hybrid descriptors\cite{Zhang_BookChap_NIPS_2018_v31_p4436,Wang_NuclFusion_2022_v62_p126013,Zhang_2022_DPA1,Zhang_JChemPhys_2022_v156_p124107},
the ability to fit tensorial properties\cite{Zhang_PhysRevB_2020_v102_p41121,Sommers_PhysChemChemPhys_2020_v22_p10592},
type embedding,
model deviation\cite{Zhang_PhysRevMater_2019_v3_p23804,Zhang_ComputPhysCommun_2020_v253_p107206},
Deep Potential - Range Correction (DPRc)\cite{Zeng_JChemTheoryComput_2021_v17_p6993,Giese_JChemTheoryComput_2022_v18_p4304},
Deep Potential Long Range (DPLR)\cite{Zhang_JChemPhys_2022_v156_p124107},
graphics processing unit (GPU) support for customized operators\cite{Lu_CompPhysCommun_2021_v259_p107624},
model compression\cite{Lu_JChemTheoryComput_2022_v18_p5559},
non-von Neumann molecular dynamics (NVNMD)\cite{Mo_npjComputMater_2022_v8_p107},
and various usability improvements such as documentation, compiled binary packages, graphical user interfaces (GUI), and application programming interfaces (API).
This article provides an overview of the current major additions to the DeePMD-kit, highlighting its features and technical details, benchmarking the accuracy and efficiency of different models, and discussing ongoing developments.

\section{Features}\label{sec:theory}

\begin{figure}
    \centering
    \includegraphics[width=\linewidth]{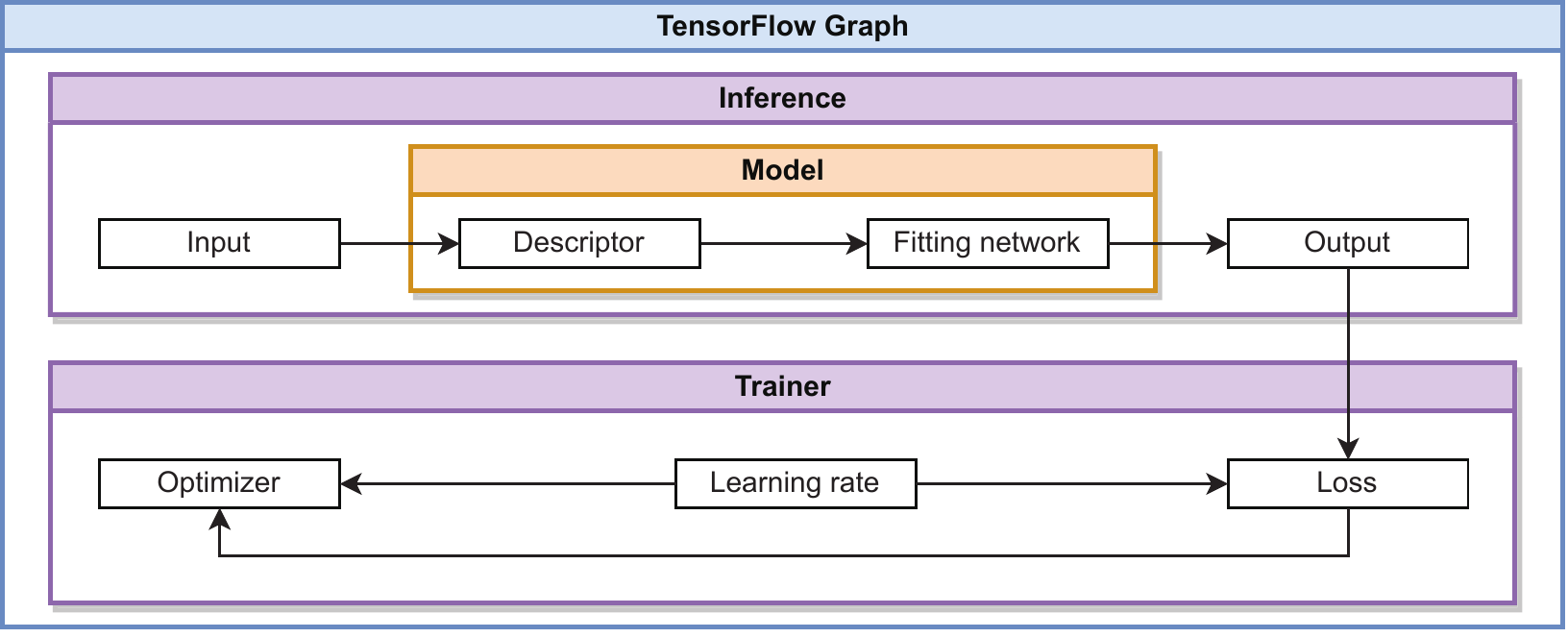}
    
    \caption{The components of the DeePMD-kit package. The direction of the arrow indicates the dependency between the components.}
    \label{fig:feature}
\end{figure}

In this section, we introduce features from the perspective of components (shown in Fig.~\ref{fig:feature}).
A component represents units of computation.
It is organized as a Python class inside the package, and a corresponding TensorFlow static graph will be created at runtime.

\subsection{Models}

A Deep Potential (DP) model, denoted by $\mathcal{M}$, can be generally represented as
\begin{align}
\bm y_i = \mathcal M (\bm x_i, \{\bm x_j\}_{j\in n(i)}; \bm \theta)
= \mathcal{F} \big( \mathcal{D} (\bm x_i, \{\bm x_j\}_{j\in n(i)}; \bm \theta_d) ; \bm \theta_f \big),
\label{eq:model}
\end{align}
where $\bm{y}_i$ is the fitting properties, $\mathcal{F}$ is the fitting network (introduced in Section \ref{section:fitting}), $\mathcal{D}$ is the descriptor (introduced in Section \ref{section:descriptor}).
$\bm{x} = (\bm r_i, \alpha_i)$, with $\bm r_i$ being the Cartesian coordinates and $\alpha_i$ being the chemical species, denotes the degrees of freedom of the atom $i$.
The indices of the neighboring atoms (i.e.~atoms within a certain cutoff radius) of atom $i$ are given by the notation $n(i)$.
Note that the Cartesian coordinates can be either under the periodic boundary condition (PBC) or in vacuum (under the open boundary condition).
The network parameters are denoted by $\bm \theta = \{\bm \theta_d, \bm \theta_f\}$, where $\bm \theta_d$ and $\bm\theta_f$ yield the network parameters of the descriptor (if any) and those of the fitting network, respectively.
From Eq.~\eqref{eq:model}, one may compute the global property of the system by
\begin{align}
    \bm y = \sum_{i=1}^N \bm y_i,
\end{align}
where $N$ is the number of atoms in a frame.
For example, if $y_i$ represents the potential energy contribution of atom $i$, then $y$ gives the total potential energy of the frame.
In the following text, $N_c$ is the expected maximum number of neighboring atoms, which is the same constant for all atoms over all frames.
A matrix with a dimension of $N_c$ will be padded if the number of neighboring atoms is less than $N_c$.

\subsubsection{Descriptors}\label{section:descriptor}

DeePMD-kit supports multiple atomic descriptors, including the local frame descriptor, two-body and three-body embedding DeepPot-SE descriptor, the attention-based descriptor, and the hybrid descriptor that is defined as a combination of multiple descriptors.
In the following text, we use $\mathcal{D}^i = \mathcal{D} (\bm x_i, \{\bm x_j\}_{j\in n(i)}; \bm \theta_d) $ to represent the atomic descriptor of the atom $i$.

\paragraph{Local frame.}
The local frame descriptor $\mathcal{D}^i \in \mathbb{R}^{N_c \times \{1,4\}}$ (sometimes simply called the DPMD model), which is the first version of the DP descriptor\cite{Zhang_PhysRevLett_2018_v120_p143001}, is constructed by using either full information or radial-only information
\begin{equation}\label{eq:descrpt-lf}
    (\mathcal{D}^i)_j = 
    \begin{cases}
    \{
    \begin{array}{cccc}
    \frac{1}{r_{ij}} & \frac{x_{ij}}{r_{ij}} & \frac{y_{ij}}{r_{ij}} & \frac{z_{ij}}{r_{ij}}
    \end{array}
    \}, &\text{full},  \\
    \{
    \begin{array}{c}
    \frac{1}{r_{ij}}
    \end{array}
    \}, &\text{radial-only}, 
    \end{cases}
\end{equation}
where $(x_{ij}, y_{ij}, z_{ij})$ are three Cartesian coordinates of the relative position between atoms $i$ and $j$, i.e.~$\bm{r}_{ij}=\bm r_i-\bm r_j=(x_{ij}, y_{ij}, z_{ij})$ in  the local frame, and $r_{ij} = \lvert \bm r_{ij} \lvert$ is its norm.
In Eq.~\eqref{eq:descrpt-lf}, the order of the neighbors $j$ is sorted in ascending order according to their distance to the atom $i$.
$\bm r_{ij}$ is transformed from the global relative coordinate $\bm r_{ij}^0$ through
\begin{equation}
    \bm r_{ij}=\bm r^0_{ij} \cdot {R}_i, \label{eq:rij_loc}
\end{equation}
where
\begin{equation}
    {R}_i=\{\mathbf{e}_{i1}, \mathbf{e}_{i2}, \mathbf{e}_{i3}\}
\end{equation}
is the rotation matrix constructed by
\begin{align}
    \bm{e}_{i1} &= \bm{e}(\bm r_{i,a(i)}), \\
    \bm{e}_{i2} &= \bm{e}(\bm r_{i,b(i)}- (\bm r_{i,b(i)}\cdot \bm{e}_{i1})\bm{e}_{i1} ), \\
    \bm{e}_{i3} &= \bm{e}_{i1} \times \bm{e}_{i2},
\end{align}
where $\bm{e}(\bm{r}_{ij}) = {\bm{r}_{ij}}/{r_{ij}}$ denotes the operation of  normalizing a vector.
$a(i)\in n(i)$ and $b(i)\in n(i)$ are the two axis atoms used to define the axes of the local frame of atom $i$, which in general, are the two closest atoms, independently of their species, together with the center atom $i$.

The limitation of the local frame descriptor is that it is not smooth at the cutoff radius and the exchanging of the order of two nearest neighbors (i.e.~the swapping of $a(i)$ and $b(i)$), so its usage is limited.
We note that the local frame descriptor is the only non-smooth descriptor among all DP descriptors, and we recommend using other descriptors for the usual system.

\paragraph{Two-body embedding DeepPot-SE.}\label{section:e2}
The two-body embedding smooth edition of the DP descriptor $\mathcal{D}^i \in \mathbb{R}^{M \times M_{<}}$, is usually named DeepPot-SE descriptor\cite{Zhang_BookChap_NIPS_2018_v31_p4436}.
It is noted that the descriptor is a multi-body representation of the local environment of the atom $i$. 
We call it ``two-body embedding'' because the embedding network takes only the distance between atoms $i$ and $j$ (see below), but it is not implied that the descriptor takes only the pairwise information between $i$ and its neighbors. 
The descriptor, using either full information or radial-only information, is given by
\begin{equation}
    \mathcal{D}^i = 
    \begin{cases}
    \frac{1}{N_c^2} (\mathcal{G}^i)^T \mathcal{R}^i (\mathcal{R}^i)^T \mathcal{G}^i_<, &\text{full}, \\
    \frac{1}{N_c} \sum_j (\mathcal{G}^i)_{jk}, &\text{radial-only}, 
    \end{cases}
\end{equation}
where $\mathcal{R}^i \in \mathbb{R}^{N_c \times \{1,4\}}$ is the coordinate matrix, and each row of $\mathcal{R}^i$ can be constructed as
\begin{equation}
    (\mathcal{R}^i)_j =
    \begin{cases}
    \{
    \begin{array}{cccc}
    s(r_{ij}) & \frac{s(r_{ij})x_{ij}}{r_{ij}} & \frac{s(r_{ij})y_{ij}}{r_{ij}} & \frac{s(r_{ij})z_{ij}}{r_{ij}} 
    \end{array}
    \}, &\text{full},  \\
    \{
    \begin{array}{c}
    s(r_{ij})
    \end{array}
    \}, &\text{radial-only}, 
    \end{cases}
    \label{eq:rij}
\end{equation}
where $\bm{r}_{ij}=\bm{r}_j-\bm{r}_i = (x_{ij}, y_{ij}, z_{ij})$ is the relative coordinate and $r_{ij}=\lVert \bm{r}_{ij} \lVert$ is its norm. The switching function $s(r)$ is defined as
\begin{equation}
    s(r)=
    \begin{cases}
    \frac{1}{r}, & r<r_s, \\
    \frac{1}{r} \big[ x^3 (-6 x^2 +15 x -10) +1 \big], & r_s \leq r<r_c, \\
    0, & r \geq r_c,
    \end{cases}
    \label{eq:switch}
\end{equation}
where $x=\frac{r - r_s}{ r_c - r_s}$  switches from 1 at $r_s$ to 0 at the cutoff radius $r_c$. 
The switching function $s(r)$ is smooth in the sense that the second-order derivative is continuous.

Each row of the embedding matrix  $\mathcal{G}^i \in \mathbb{R}^{N_c \times M}$ consists of $M$ nodes from the output layer of an NN function $\mathcal{N}_g$ of $s(r_{ij})$:
\begin{equation}
    (\mathcal{G}^i)_j = \mathcal{N}_{e,2}(s(r_{ij})),
    \label{eq:G2}
\end{equation}
where the NN function will be introduced in Section \ref{section:NN}, and the subscript ``$e,2$'' is used to distinguish the NN from other NNs used in the DP model.
In Eq.~\eqref{eq:G2}, the network parameters are not explicitly written.
$\mathcal{G}^i_< \in \mathbb{R}^{N_c \times M_<}$ only takes first $M_<$ columns of $\mathcal{G}^i$ to reduce the size of $\mathcal D^i$.
$r_s$, $r_c$, $M$ and $M_<$ are hyperparameters provided by the user.
Compared to the local frame descriptor, the DeepPot-SE is continuous up to the second-order derivative in its domain.

\paragraph{Three-body embedding DeepPot-SE.}\label{section:e3}
The three-body embedding DeepPot-SE descriptor incorporates bond-angle information, making the model more accurate\cite{Wang_NuclFusion_2022_v62_p126013}. The descriptor $\mathcal{D}^i$ can be represented as
\begin{equation}\label{eq:descrpt-e3}
    \mathcal{D}^i = \frac{1}{N_c^2}(\mathcal{R}^i(\mathcal{R}^i)^T):\mathcal{G}^i,
\end{equation}
where $\mathcal{R}^i$ is defined by Eq.~\eqref{eq:rij}. 
Currently, only the full information case of $\mathcal{R}^i$ is supported by the three-body embedding.
Similar to Eq.~\eqref{eq:G2}, each element of $\mathcal{G}^i \in \mathbb{R}^{N_c \times N_c \times M}$ comes from $M$ nodes from the output layer of an NN $\mathcal{N}_{e,3}$ function:
\begin{equation}\label{eq:nn-descrpt-e3}
    (\mathcal{G}^i)_{jk}=\mathcal{N}_{e,3}((\theta_i)_{jk}),
\end{equation}
where $(\theta_i)_{jk} = (\mathcal{R}^i)_j\cdot (\mathcal{R}^i)_k$ considers the angle form of two neighbours ($j$ and $k$).
The notation ``$:$'' in Eq.~\eqref{eq:descrpt-e3} indicates the contraction between matrix $\mathcal{R}^i(\mathcal{R}^i)^T$ and the first two dimensions of tensor $\mathcal{G}^i$.
The network parameters are also not explicitly written in Eq.~\eqref{eq:nn-descrpt-e3}.

\paragraph{Handling the systems composed of multiple chemical species.}
For a system with multiple chemical species ($|\{\alpha_i\}| > 1$), parameters of the embedding network $\mathcal{N}_{e,\{2,3\}}$ are as follows chemical-species-wise in Eqs.~\eqref{eq:G2} and \eqref{eq:nn-descrpt-e3}:
\begin{align}
    &(\mathcal{G}^i)_j = \mathcal{N}^{\alpha_i, \alpha_j}_{e,2}(s(r_{ij})) \quad \mathrm{or}\quad
    (\mathcal{G}^i)_j = \mathcal{N}^{ \alpha_j}_{e,2}(s(r_{ij})),
    \\
    &(\mathcal{G}^i)_{jk} =\mathcal{N}^{\alpha_j, \alpha_k}_{e,3}((\theta_i)_{jk}).
\end{align}
Thus, there will be $N_t$ or $N_t^2$ embedding networks where $N_t$ is the number of chemical species.
To improve the performance of matrix operations, $n(i)$ is divided into blocks of different chemical species.
Each matrix with a dimension of $N_c$ is divided into corresponding blocks, and each block is padded to $N_c^{\alpha_j}$ separately.
The limitation of this approach is that when there are large numbers of chemical species, the number of embedding networks will explode, requiring large memory and decreasing computing efficiency.

\paragraph{Type embedding.}\label{section:tebd}

To reduce the number of NN parameters and improve computing efficiency when there are large numbers of chemical species,
the type embedding $\mathcal{A}$ is introduced, represented as a NN function $\mathcal{N}_t$ of the atomic type $\alpha$:
\begin{equation}
    \mathcal{A}^i = \mathcal{N}_t\big( \text{one\textunderscore hot}(\alpha_i) \big),
\end{equation}
where $\alpha_i$ is converted to a one-hot vector representing the chemical species before feeding to the NN.
The NN function will be introduced in Section~\ref{section:NN}.
Based on Eqs.~\eqref{eq:G2} and \eqref{eq:nn-descrpt-e3}, the type embeddings of central and neighboring atoms $\mathcal{A}^i$ and $\mathcal{A}^j$ are added as an extra input of the embedding network $\mathcal{N}_{e,\{2,3\}}$:
\begin{align}
    &(\mathcal{G}^i)_j = \mathcal{N}_{e,2}(\{s(r_{ij}), \mathcal{A}^i, \mathcal{A}^j\})  \quad \mathrm{or}\quad
    (\mathcal{G}^i)_j = \mathcal{N}_{e,2}(\{s(r_{ij}), \mathcal{A}^j\}) , \label{eq:G2_tbed} \\
    &(\mathcal{G}^i)_{jk} =\mathcal{N}_{e,3}(\{(\theta_i)_{jk}, \mathcal{A}^j, \mathcal{A}^k\}).
\end{align}
In this way, all chemical species share the same network parameters through the type embedding.

\paragraph{Attention-based descriptor.}
Attention-based descriptor $\mathcal{D}^i \in \mathbb{R}^{M \times M_{<}}$, which is proposed in pretrainable DPA-1 \cite{Zhang_2022_DPA1} model, is given by
\begin{equation}
    \mathcal{D}^i = \frac{1}{N_c^2}(\hat{\mathcal{G}}^i)^T \mathcal{R}^i (\mathcal{R}^i)^T \hat{\mathcal{G}}^i_<,
\end{equation}
where $\hat{\mathcal{G}}^i$ represents the embedding matrix $\mathcal{G}^i$ after additional self-attention mechanism\cite{Vaswani_2017_Attention} and $\mathcal{R}^i$ is defined by the full case in the Eq.~\eqref{eq:rij}. 
Note that we obtain $\mathcal{G}^i$ from Eq.~\eqref{eq:G2_tbed} using the type embedding method by default in this descriptor. 

To perform the self-attention mechanism, the queries $\mathcal{Q}^{i,l} \in \mathbb{R}^{N_c\times d_k}$, keys $\mathcal{K}^{i,l} \in \mathbb{R}^{N_c\times d_k}$, and values $\mathcal{V}^{i,l} \in \mathbb{R}^{N_c\times d_v}$ are first obtained:
\begin{eqnarray}
    \left(\mathcal{Q}^{i,l}\right)_{j}&=Q_{l}\left(\left(\mathcal{G}^{i,l-1}\right)_{j}\right),\\
    \left(\mathcal{K}^{i,l}\right)_{j}&=K_{l}\left(\left(\mathcal{G}^{i,l-1}\right)_{j}\right), \\
    \left(\mathcal{V}^{i,l}\right)_{j}&=V_{l}\left(\left(\mathcal{G}^{i,l-1}\right)_{j}\right),
\end{eqnarray}
where $Q_{l}$, $K_{l}$, $V_{l}$ represent three trainable linear transformations that output the queries and keys of dimension $d_k$ and values of dimension $d_v$, and $l$ is the index of the attention layer.
The input embedding matrix to the attention layers,  denoted by $\mathcal{G}^{i,0}$, is chosen as the two-body embedding matrix~\eqref{eq:G2}.

Then the scaled dot-product attention method\cite{NIPS2017_3f5ee243,luong2015effective} is adopted:
\begin{equation}
A(\mathcal{Q}^{i,l}, \mathcal{K}^{i,l}, \mathcal{V}^{i,l}, \mathcal{R}^{i,l})=\varphi\left(\mathcal{Q}^{i,l}, \mathcal{K}^{i,l},\mathcal{R}^{i,l}\right)\mathcal{V}^{i,l},
\end{equation}
where $\varphi\left(\mathcal{Q}^{i,l}, \mathcal{K}^{i,l},\mathcal{R}^{i,l}\right) \in \mathbb{R}^{N_c\times N_c}$ is attention weights.
In the original attention method, one typically has $\varphi\left(\mathcal{Q}^{i,l}, \mathcal{K}^{i,l}\right)=\operatorname{softmax}\left(\frac{\mathcal{Q}^{i,l} (\mathcal{K}^{i,l})^{T}}{\sqrt{d_{k}}}\right)$, with $\sqrt{d_{k}}$ being the normalization temperature.
This is slightly modified to incorporate the angular information:
\begin{equation}
\varphi\left(\mathcal{Q}^{i,l}, \mathcal{K}^{i,l},\mathcal{R}^{i,l}\right) = \operatorname{softmax}\left(\frac{\mathcal{Q}^{i,l} (\mathcal{K}^{i,l})^{T}}{\sqrt{d_{k}}}\right) \odot \hat{\mathcal{R}}^{i}(\hat{\mathcal{R}}^{i})^{T},
\end{equation}
where $\hat{\mathcal{R}}^{i} \in \mathbb{R}^{N_c\times 3}$ denotes normalized relative coordinates , $\hat{\mathcal{R}}^{i}_{j} = \frac{\bm{r}_{ij}}{\lVert \bm{r}_{ij} \lVert}$ and $\odot$ means element-wise multiplication. 

Then layer normalization is added in a residual way to finally obtain the self-attention local embedding matrix $\hat{\mathcal{G}}^{i} = \mathcal{G}^{i,L_a}$ after $L_a$  attention layers:
\begin{equation}
\mathcal{G}^{i,l} = \mathcal{G}^{i,l-1} + \mathrm{LayerNorm}(A(\mathcal{Q}^{i,l}, \mathcal{K}^{i,l}, \mathcal{V}^{i,l}, \mathcal{R}^{i,l})).
\end{equation}

\paragraph{Hybrid descriptor.}
A hybrid descriptor $\mathcal{D}^i_\text{hyb}$ concatenates multiple kinds of descriptors into one descriptor:\cite{Zhang_JChemPhys_2022_v156_p124107}
\begin{equation}
    \mathcal{D}^{i}_\text{hyb} = \{
    \begin{array}{cccc}
        \mathcal{D}^{i}_1 & \mathcal{D}^{i}_2 & \cdots & \mathcal{D}^{i}_n
    \end{array}
    \}.
\end{equation}
The list of descriptors can be different types or the same descriptors with different parameters.
This way, one can set the different cutoff radii for different descriptors.

\subsubsection{Fitting networks}\label{section:fitting}
The fitting network can fit the potential energy of a system, along with the force and the virial, and tensorial properties such as the dipole and the polarizability.

\paragraph{Fitting potential energies.}
In the DP model~\eqref{eq:model}, we let the fitting network $\mathcal{F}_0$ maps the descriptor $\mathcal D^i$ to a scalar, where the subscript ``0'' means that the output is a zero-order tensor (i.e.~scalar).  The model can then be used to predict the total potential energy of the system by
\begin{align}\label{eq:fitting_energy}
    E  =  \sum_i E_i = \sum_i \mathcal F_0 (\mathcal D^i),
\end{align}
where the output of the fitting network is treated as the atomic potential energy contribution, i.e.~$E_i$. 
The output scalar can also be treated as other scalar properties defined on an atom, for example, the partial charge of atom $i$.

In some cases, atomic-specific or frame-specific  parameters, such as electron temperature\cite{Zhang_PhysPlas_2020_v27_p122704}, may be treated as extra input to the fitting network. 
We denote the atomic and frame-specific parameters by $\bm{P}^i\in \mathbb{R}^{N_p}$ (with $N_p$ being the dimension) and $\bm{Q}\in \mathbb{R}^{N_q}$ (with $N_q$ being the dimension), respectively. 
\begin{equation}\label{eq:fitting_network}
    E_i=\mathcal{F}_0(\{\mathcal{D}^i, \bm{P}^i, \bm Q\}).
\end{equation}

The atomic force $\bm{F}_i$ and the virial tensor $\bm{\Xi} = (\Xi_{\alpha\beta})$ (if PBC is applied) can be derived from the potential energy $E$:
\begin{align}
    F_{i,\alpha}&=-\frac{\partial E}{\partial r_{i,\alpha}}, \\
    \Xi_{\alpha\beta}&=-\sum_{\gamma} \frac{\partial E}{\partial h_{\gamma\alpha}} h_{\gamma\beta},
\end{align}
where $r_{i,\alpha}$ and $F_{i,\alpha}$ denotes the $\alpha$-th component of the coordinate and force of atom $i$. $h_{\alpha\beta}$ is the $\beta$-th component of the $\alpha$-th basis vector of the simulation region. 

\paragraph{Fitting tensorial properties.}
To represent the first-order tensorial properties (i.e.~vector properties), we let the fitting network, denoted by $\mathcal F_{1}$, output an $M$-dimensional vector; then we have the representation,
\begin{align}
(T_i^{(1)})_\alpha = 
\frac{1}{N_c}
\sum_{j=1}^{N_c}\sum_{m=1}^M (\mathcal G^i)_{jm} (\mathcal R^i)_{j,\alpha+1} 
(\mathcal F_{1}(\mathcal D^i))_m, \ \alpha=1,2,3.
\end{align}
We let the fitting network $\mathcal F_{2}$ output an $M$-dimensional vector, and the second-order tensorial properties (matrix properties) are formulated as
\begin{align}\label{eq:dipole}
(T_i^{(2)})_{\alpha\beta} = 
\frac{1}{N_c^2}
\sum_{j=1}^{N_c}\sum_{k=1}^{N_c}\sum_{m=1}^M 
(\mathcal G^i)_{jm}
(\mathcal R^i)_{j,\alpha+1} 
(\mathcal R^i)_{k,\beta+1} 
(\mathcal G^i)_{km}
(\mathcal F_{2}(\mathcal D^i))_m, 
\ \alpha,\beta=1,2,3,
\end{align}
where $\mathcal{G}^i$ and $\mathcal{R}^i$ can be found at Eq.~\eqref{eq:G2} and \eqref{eq:rij} (full case), respectively.
Thus, the tensor fitting network requires the descriptor to have the same or similar form as the DeepPot-SE descriptor.
The NN functions $\mathcal{F}_1$ and $\mathcal F_2$ will be introduced in Section~\ref{section:NN}.
The total tensor $\bm{T}$ (total dipole $\bm{T}^{(1)}$ or total polarizability $\bm{T}^{(2)}$) is the sum of the atomic tensor:
\begin{equation}\label{eq:total_tensor}
    \bm{T} = \sum_i \bm{T}_i.
\end{equation}
The tensorial models can be used to calculate IR spectrum\cite{Zhang_PhysRevB_2020_v102_p41121} and Raman spectrum\cite{Sommers_PhysChemChemPhys_2020_v22_p10592}.

\paragraph{Handling the systems composed of multiple chemical species.}
Similar to the embedding networks, if the type embedding approach is not used, the fitting network parameters are chemical-species-wise, and there are $N_t$ sets of fitting network parameters.
For performance, atoms are sorted by their chemical species $\alpha_i$ in advance.
Take an example, the atomic energy $E_i$ is represented as follows based on Eq.~\eqref{eq:fitting_network}:
\begin{equation}
E_i=\mathcal{F}_0^{\alpha_i}(\mathcal{D}^i).
\end{equation}
When the type embedding is used, all chemical species share the same network parameters, and the type embedding is inserted into the input of the fitting networks in Eq.~\eqref{eq:fitting_network}:
\begin{equation}
E_i=\mathcal{F}_0(\{\mathcal{D}^i, \mathcal{A}^i\}).
\end{equation}

\subsubsection{Deep Potential Range Correction (DPRc)}

Deep Potential - Range Correction (DPRc)\cite{Zeng_JChemTheoryComput_2021_v17_p6993,Giese_JChemTheoryComput_2022_v18_p4304} was initially designed to correct the potential energy from a fast, linear-scaling low-level semiempirical QM/MM theory to a high-level \textit{ab initio} QM/MM theory in a range-correction way to quantitatively correct short and mid-range non-bonded interactions leveraging the non-bonded lists routinely used in molecular dynamics simulations using molecular mechanical force fields such as AMBER.\cite{Lee_JChemInfModel_2018_v58_p2043} 
In this way, long-ranged electrostatic interactions can be modeled efficiently using the particle mesh Ewald method\cite{Lee_JChemInfModel_2018_v58_p2043} or its extensions for multipolar\cite{Giese_JChemTheoryComput_2015_v11_p436,Giese_JChemTheoryComput_2015_v11_p451} and QM/MM\cite{Nam_JChemTheoryComput_2005_v1_p2,Giese_JChemTheoryComput_2016_v12_p2611} potentials.
In a DPRc model, the switch function in Eq.~\eqref{eq:switch} is modified to disable MM-MM interaction:
\begin{equation}\label{E:s2}
  s_\text{DPRc}(r_{ij}) =
  \begin{cases}
  0, &\text{if $i \in \text{MM} \land j \in \text{MM}$}, \\
  s(r_{ij}), &\text{otherwise},
  \end{cases}
\end{equation}
where $s_\text{DPRc}(r_{ij})$ is the new switch function and $s(r_{ij})$ is the old one in Eq.~\eqref{eq:switch}.
This ensures the forces between MM atoms are zero, i.e. 
\begin{equation}
{\bm F}_{ij} = - \frac{\partial E}{\partial \bm r_{ij}} = 0, \quad i \in \text{MM} \land j \in \text{MM}.
\end{equation}
The fitting network in Eq.~\eqref{eq:fitting_network} is revised to remove energy bias from MM atoms:
\begin{equation}
  E_i=
  \begin{cases}
  \mathcal{F}_0(\mathcal{D}^i),  &\text{if $i \in \text{QM}$}, \\
  \mathcal{F}_0(\mathcal{D}^i) - \mathcal{F}_0(\mathbf{0}), &\text{if $i \in \text{MM}$},
  \end{cases}
\end{equation}
where $\mathbf{0}$ is a zero matrix.
It is worth mentioning that usage of DPRc is not limited to its initial design for QM/MM correction and can be expanded to any similar interaction\cite{yang2023new}.

\subsubsection{Deep Potential Long Range (DPLR)}

The Deep Potential Long Range (DPLR) model adds the electrostatic energy to the total energy\cite{Zhang_JChemPhys_2022_v156_p124107}:
\begin{equation}
    E=E_{\text{DP}} + E_{\text{ele}},
\end{equation}
where $E_{\text{DP}}$ is the short-range contribution constructed as the standard energy model in Eq.~\eqref{eq:fitting_energy} that is fitted against $(E^\ast-E_{\text{ele}})$. 
$E_{\text{ele}}$ is the electrostatic energy
introduced by a group of Gaussian distributions that is an approximation of the electronic structure of the system, and is calculated in Fourier space by
\begin{equation}
    E_{\text{ele}} = \frac{1}{2\pi V}\sum_{m \neq 0, \|m\|\leq L} \frac{\exp({-\pi ^2 m^2/\beta ^2})}{m^2}S^2(m),
\end{equation}
where $\beta$ is a freely tunable parameter that controls the spread of the Gaussians. 
$L$ is the cutoff in Fourier space and $S(m)$, the structure factor, is given by
\begin{equation}
    S(m)=\sum_i q_i e^{-2\pi \imath m \bm r_i} + \sum_n q_n e^{-2\pi \imath m \bm W_n},
\end{equation}
where $\imath = \sqrt{-1}$ denotes the imaginary unit, $\bm r_i$ indicates ion coordinates, $q_i$ is the charge of the ion $i$, and $W_n$ is the $n$-th Wannier centroid (WC) which can be obtained from a separated dipole model in Eq.~\eqref{eq:dipole}.
It can be proved that the error in the electrostatic energy introduced by the Gaussian approximations is dominated by a summation of dipole-quadrupole interactions that decay as $r^{-4}$, where $r$ is the distance between the dipole and quadrupole~\cite{Zhang_JChemPhys_2022_v156_p124107}.

\subsubsection{Interpolation with a pairwise potential}

In applications like the radiation damage simulation, the interatomic distance may become too close, so that the DFT calculations fail. 
In such cases, the DP model that is an approximation of the DFT potential energy surface is usually replaced by an empirical potential, like the Ziegler-Biersack-Littmark (ZBL)~\cite{Ziegler_BookChap_1985_ZBL} screened nuclear repulsion potential in the radiation damage simulations~\cite{Wang_ApplPhysLett_2019_v114_p244101}.
The DeePMD-kit package supports the interpolation between DP and an empirical pairwise potential 
\begin{align}
  E_i = (1-w_i) E_i^{\mathrm{DP}} + w_i E_i^{\mathrm{pair}},
\end{align}
where the $w_i$ is the interpolation weight and the $E_i^{\mathrm{pair}}  $ is the atomic contribution due to the pairwise potential $u^{\mathrm{pair}}(r)$, i.e.
\begin{align}
  E_i^{\mathrm{pair}} = \sum_{j\in n(i)} u^{\mathrm{pair}}(r_{ij}).
\end{align}
The interpolation weight $w_i$ is defined by
\begin{equation}
    w_i =
    \begin{cases}
    1, & \sigma_i <r_a, \\
    u_i^3 (-6 u_i^2 +15 u_i -10) +1, & r_a \leq \sigma_i <r_b, \\
    0, & \sigma_i \geq r_b,
    \end{cases}
    \label{eq:interpolation}
\end{equation}
where $u_i = (\sigma_i - r_a ) / (r_b - r_a)$. In the range $[r_a, r_b]$, the DP model smoothly switched off and the pairwise potential smoothly switched on from $r_b$ to $r_a$. The $\sigma_i$ is the softmin of the distance between atom $i$ and its neighbors,
\begin{align}
  \sigma_i =
  \dfrac
  {\sum\limits_{j\in n(i)} r_{ij} e^{-r_{ij} / \alpha_s}}
  {\sum\limits_{j\in n(i)} e^{-r_{ij} / \alpha_s}},
\end{align}
where the scale $\alpha_s$ is a tunable scale of the interatomic distance $r_{ij}$.
The pairwise potential $u^{\textrm{pair}}(r)$ is defined by a user-defined table that provides the value of $u^{\textrm{pair}}$ on an evenly discretized grid from 0 to the cutoff distance.

\subsubsection{Neural networks}\label{section:NN}
\paragraph{Neural networks.}
A neural network (NN) function $\mathcal{N}$ is the composition of multiple layers $\mathcal{L}^{(i)}$:
\begin{equation}
    \mathcal{N} = \mathcal{L}^{(n)} \circ \mathcal{L}^{(n-1)} \circ \cdots \circ \mathcal{L}^{(1)}.
\end{equation}
In the DeePMD-kit package, a layer $\mathcal{L}$ may be one of the following forms, depending on whether a ResNet\cite{He_BookCharp_ECCV_2016_p630} is used and the number of nodes:
\begin{equation}
    \bm{y}=\mathcal{L}(\bm{x};\bm{w},\bm{b})=
    \begin{cases}
        \bm{\hat{w}} \odot \boldsymbol{\phi}(\bm{x}^T\bm{w}+\bm{b}) + \bm{x}, & \text{ResNet and } N_2=N_1, \\
        \bm{\hat{w}} \odot \boldsymbol{\phi}(\bm{x}^T\bm{w}+\bm{b}) + \{\bm{x}, \bm{x}\}, & \text{ResNet and } N_2 = 2N_1,\\
        \bm{\hat{w}} \odot \boldsymbol{\phi}(\bm{x}^T\bm{w}+\bm{b}), & \text{otherwise}, \\
    \end{cases}
    \label{eq:NN}
\end{equation}
where $\bm{x} \in \mathbb{R}^{N_1}$ is the input vector and $\bm{y} \in \mathbb{R}^{N_2}$ is the output vector.
$\bm{w} \in \mathbb{R}^{N_1 \times N_2}$ and $\bm{b} \in \mathbb{R}^{N_2}$ are weights and biases, respectively, both of which are trainable.
$\bm{\hat{w}} \in \mathbb{R}^{N_2}$ can be either a trainable vector, which represents the ``timestep'' in the skip connection, or a vector of all ones $\mathbf{1} = \{1, 1, \dots, 1\}$, which disables the timestep.
$\boldsymbol{\phi}$ is the activation function.
In theory, the activation function can be any form, and the following functions are provided in the DeePMD-kit package: hyperbolic tangent (tanh), rectified linear unit (ReLU)\cite{relu}, ReLU6, softplus\cite{softplus}, sigmoid, Gaussian error linear unit (GELU)\cite{hendrycks2020gaussian}, and identity.
Among these activation functions, ReLU and ReLU6 are not continuous in the first-order derivative, and others are continuous up to the second-order derivative.

\paragraph{Compression of neural networks.}
The compression of the DP model uses three techniques, tabulated inference, operator merging, and precise neighbor indexing, to improve the performance of model training and inference when the model parameters are properly trained~\cite{Lu_JChemTheoryComput_2022_v18_p5559}.

For better  performance, the NN inference can be replaced by tabulated function evaluations if the input of the NN is of dimension one. 
The embedding networks $\mathcal N_{e,2}$ defined by \eqref{eq:G2} and $\mathcal N_{e,3}$ defined by \eqref{eq:nn-descrpt-e3}  are of this type.
The idea is to approximate the output of the NN by a piece-wise polynomial fitting.
The input domain (a compact domain in $\mathbb R$) is divided into $L_c$ equally spaced intervals, in which apply a fifth-order polynomial $g^l_m(x)$ approximation of the $m$-th output component of the NN function: 
\begin{equation}
    g^l_m(x) = a^l_m x^5 + b^l_m x^4 + c^l_m x^3 + d^l_m x^2 + e^l_m x + f^l_m,\quad
    x \in [x_l, x_{l+1}),
    \label{eq:compress}
\end{equation}
where $l=1,2,\dots,L_c$ is the index of the intervals, $x_1, \dots, x_{L_c}, x_{L_c+1}$ are the endpoints of the intervals, and $a^l_m$, $b^l_m$, $c^l_m$, $d^l_m$, $e^l_m$, and $f^l_m$ are the fitting parameters.
The fitting parameters can be computed by the equations below:
\begin{align}
    a^l_m &= \frac{1}{2\Delta x_l^5}[12h_{m,l}-6(y'_{m,l+1}+y'_{m,l})\Delta x_l + (y''_{m,l+1}-y''_{m,l})\Delta x_l^2], \\
    b^l_m &= \frac{1}{2\Delta x_l^4}[-30h_{m,l} +(14y'_{m,l+1}+16y'_{m,l})\Delta x_l + (-2y''_{m,l+1}+3y''_{m,l})\Delta x_l^2], \\
    c^l_m &= \frac{1}{2\Delta x_l^3}[20h_{m,l}-(8y'_{m,l+1}+12y'_{m,l})\Delta x_l + (y''_{m,l+1}-3y''_{m,l})\Delta x_l^2], \\
    d^l_m &= \frac{1}{2}y''_{m,l}, \\
    e^l_m &= y_{m,l}', \\
    f^l_m &= y_{m,l}, 
\end{align}
where $\Delta x_l=x_{l+1}-x_l$ denotes the size of the interval. $h_{m,l}=y_{m,l+1}-y_{m,l}$. $y_{m,l} = y_m(x_l)$, $y'_{m,l} = y'_m(x_l)$ and $y''_{m,l} = y''_m(x_l)$ are the value, the first-order derivative, and the second-order derivative of the $m$-th component of the target NN function at the interval point $x_l$, respectively.
The first and second-order derivatives are easily calculated by the back-propagation of the NN functions. 

In the standard DP model inference, taking the two-body embedding descriptor as an example, the matrix product $(\mathcal G^i)^T \mathcal R$ requires the transfer of the tensor  $\mathcal G^i$ between the register and the host/device memories, which usually becomes the bottle-neck of the computation due to the relatively small memory bandwidth of the GPUs. 
The compressed DP model merges the matrix multiplication $(\mathcal G^i)^T \mathcal R$ with the tabulated inference step.
More specifically, once one column of the $(\mathcal G^i)^T $ is evaluated, it is immediately multiplied with one row of the environment matrix in the register, and the outer product is deposited to the result of $(\mathcal G^i)^T \mathcal R$. 
By the operator merging technique, the allocation of  $\mathcal G^i$ and the memory movement between register and host/device memories is avoided. 
The operator merging of the three-body embedding can be derived analogously.

The first dimension, $N_c$, of the environment ($\mathcal R^i$) and embedding ($\mathcal G^i$) matrices is the expected maximum number of neighbors. 
If the number of neighbors of an atom is smaller than $N_c$, the corresponding positions of the matrices are pad with zeros.
In practice, if the real number of neighbors is significantly smaller than $N_c$, a notable operation is spent on the multiplication of padding zeros. 
In the compressed DP model, the number of neighbors is precisely indexed at the tabulated inference stage, further saving computational costs.

\subsection{Trainer}
Based on DP models $\mathcal{M}$ defined in Eq.~\eqref{eq:model}, a trainer should also be defined to train parameters in the model, including weights and biases in Eq.~\eqref{eq:NN}.
The learning rate $\gamma$, the loss function $L$, and the training process should be given in a trainer.

\subsubsection{Learning rate}
The learning rate $\gamma$ decays exponentially:
\begin{equation}
    \gamma(\tau) = \gamma^0 r ^ {\lfloor  \tau/s \rfloor},\label{eq:learning_rate}
\end{equation}
where $\tau \in \mathbb{N}$ is the index of the training step, $\gamma^0  \in \mathbb{R}$ is the learning rate at the first step, and the decay rate $r$ is given by
\begin{equation}
    r = {\left(\frac{\gamma^{\text{stop}}}{\gamma^0}\right )} ^{\frac{s}{\tau^{\text{stop}}}},
\end{equation}
where $\tau^{\text{stop}} \in \mathbb{N}$, $\gamma^{\text{stop}} \in \mathbb{R}$, and $s \in \mathbb{N}$ are the stopping step, the stopping learning rate, and the decay steps, respectively, all of which are hyperparameters provided in advance.

\subsubsection{Loss function}
The loss function $L$ is given by a weighted sum of different fitting property loss $L_p$:
\begin{equation}
    L(\bm{x};\boldsymbol{\theta},\tau)=\frac{1}{\mathcal{B}}\sum_{k\in\mathcal B} \sum_\eta p_\eta(\tau) L_{\eta}(\bm{x}^k;\boldsymbol{\theta}) \label{eq:loss},
\end{equation}
where $\mathcal{B}$ is the mini-batch of data.
$\bm x = \{\bm x^k\}$ is the dataset. $\bm x^k =(\bm x^k_1, \dots, \bm x^k_N )$ is a single data frame from the set and is composed of all the degrees of freedom of the atoms. 
$\eta$ denotes the property to be fit.
For each property, $p_\eta$ is a prefactor given by
\begin{equation}
    p_\eta(\tau) = p_\eta^{\text{limit}} (1-\frac{\gamma(\tau)}{\gamma^0}) + p_\eta^{\text{start}} \frac{\gamma(\tau)}{\gamma^0},
\end{equation}
where $p_\eta^{\text{start}}$ and $p_\eta^{\text{limit}}$ are hyperparameters that give the prefactor at the first training step and the infinite training steps, respectively. 
$\gamma(\tau)$ is the learning rate defined by Eq.~\eqref{eq:learning_rate}.

The loss function of a specific fitting property $L_\eta$ is defined by the mean squared error (MSE) of a data frame and is normalized by the number of atoms $N$ if $\eta$ is a frame property that is a linear combination of atomic properties.
Take an example, if an energy model is fitted as given in Eq.~\eqref{eq:fitting_energy}, the properties $\eta$ could be energy $E$, force $\bm{F}$, virial $\bm{\Xi}$, relative energy $\Delta E$\cite{Zeng_JChemTheoryComput_2023_v19_p1261}, or any combination among them, and the loss functions of them are
\begin{align}
    L_E(\bm{x};\boldsymbol{\theta})&=\frac{1}{N}(E(\bm{x};\boldsymbol{\theta})-E^*)^2, \\
    L_F(\bm{x};\boldsymbol{\theta})&=\frac{1}{3N}\sum_{k=1}^{N}\sum_{\alpha=1}^3(F_{k,\alpha}(\bm{x};\boldsymbol{\theta})-F_{k,\alpha}^*)^2, \label{eq:loss_f} \\
    L_\Xi(\bm{x};\boldsymbol{\theta})&=\frac{1}{9N}\sum_{\alpha,\beta=1}^{3}(\Xi_{\alpha\beta}(\bm{x};\boldsymbol{\theta})-\Xi_{\alpha\beta}^*)^2, \\
    L_{\Delta E}(\bm{x};\boldsymbol{\theta})&=\frac{1}{N}({\Delta E}(\bm{x};\boldsymbol{\theta})-{\Delta E}^*)^2,
\end{align}
where $F_{k,\alpha}$ is the $\alpha$-th component of the force on atom $k$, and the superscript ``$\ast$'' indicates the label of the property that should be provided in advance. 
Using $N$ ensures that each loss of fitting property is averaged over atomic contributions before they contribute to the total loss by weight.

If part of atoms is more important than others, the MSE of atomic forces with prefactors $q_{k}$ can also be used as the loss function:
\begin{align}
    L_F^p(\mathbf{x};\boldsymbol{\theta})&=\frac{1}{3N}\sum_{k=1}^{N} \sum_{\alpha} q_{k} (F_{k,\alpha}(\mathbf{x};\boldsymbol{\theta})-F_{k,\alpha}^*)^2.
\end{align}
If some forces are quite large, one may also prefer the force loss is relative to the magnitude instead of Eq.~\eqref{eq:loss_f}:
\begin{equation}
    L^r_F(\bm{x};\boldsymbol{\theta})=\frac{1}{3N}\sum_{k=1}^{N}\sum_\alpha \left(\frac{F_{k,\alpha}(\bm{x};\boldsymbol{\theta})-F_{k,\alpha}^*}{\lvert\bm{F}^\ast_k\lvert}\right)^2.
\end{equation}

\subsubsection{Training process}
During the training process, the loss function is minimized by the stochastic gradient descent algorithm Adam~\cite{Kingma_ICLR_2015}. Ideally, the resulting parameter is the minimizer of the loss function, 
\begin{align}
    \bm \theta^\ast = \underset{\bm \theta}{\mathrm{argmin}} \lim_{\tau \to +\infty} L (\bm x; \bm \theta,\tau).
\end{align}
In practice, the Adam optimizer stops at the step $\tau_{\text{stop}}$, and the learning rate varies according to the scheme~\eqref{eq:learning_rate}.
$\tau_{\text{stop}}$ is a hyperparameter usually set to several million.

\subsubsection{Multiple tasks training}
The multi-task training process can simultaneously handle different datasets with properties that can not be fitted in one network (e.g. properties from DFT calculations under different exchange-correlation functionals or different basis sets). 
These datasets are denoted by $\bm x^{(1)}, \dots, \bm x^{(n_t)}$.
For each dataset, a training task is defined as 
\begin{align}
    \min_{\bm \theta}   L^{(t)} (\bm x^{(t)}; \bm  \theta^{(t)}, \tau), \quad t=1, \dots, n_t.
\end{align}
During the multi-task training process, all tasks share one descriptor with trainable parameters $\bm{\theta}_{d}$, while each of them has its own fitting network with trainable parameters $\bm{\theta}_f^{(t)}$, thus 
$\bm{\theta}^{(t)} = \{ \bm{\theta}_{d} , \bm{\theta}_{f}^{(t)} \}$.
At each training step, a task is randomly picked from ${1, \dots, n_t}$, and the Adam optimizer is executed to minimize $L^{(t)}$ for one step to update the parameter $\bm \theta^{(t)}$.
If different fitting networks have the same architecture, they can share the parameters of some layers 
to improve training efficiency.

\subsection{Model deviation}
Model deviation $\epsilon_y$ is the standard deviation of properties $\bm y$ inferred by an ensemble of models $\mathcal{M}_1, \dots, \mathcal{M}_{n_m}$ that are trained by the same dataset(s) with the model parameters initialized independently. 
The DeePMD-kit supports $\bm y$ to be the atomic force $\bm F_i$ and the virial tensor $\bm \Xi$.
The model deviation is used to estimate the error of a model at a certain data frame, denoted by $\bm x$, containing the coordinates and chemical species of all atoms. 
We present the model deviation of the atomic force and the virial tensor
\begin{align}
    \epsilon_{\bm{F},i} (\bm x)&=
    \sqrt{\langle \lVert \bm F_i(\bm x; \bm \theta_k)-\langle \bm F_i(\bm x; \bm \theta_k) \rangle \rVert^2 \rangle}, \\
    \epsilon_{\bm{\Xi},{\alpha \beta}} (\bm x)&=
    \frac{1}{N} \sqrt{\langle ( {\Xi}_{\alpha \beta}(\bm x; \bm \theta_k)-\langle {\Xi}_{\alpha \beta}(\bm x; \bm \theta_k) \rangle )^2 \rangle},
\end{align}
where $\bm \theta_k$ is the parameters of the model $\mathcal M_k$, and the ensemble average $\langle\cdot\rangle$ is estimated by
\begin{align}
    \langle \bm y(\bm x; \bm \theta_k) \rangle 
    =
    \frac{1}{n_m} \sum_{k=1}^{n_m} \bm y(\bm x; \bm \theta_k).
\end{align}
Small $\epsilon_{\bm{F},i}$ means the model has learned the given data; otherwise, it is not covered, and the training data needs to be expanded.
If the magnitude of $\bm F_i$ or $\bm \Xi$ is quite large,
a relative model deviation $\epsilon_{\bm{F},i,\text{rel}}$ or $\epsilon_{\bm{\Xi},\alpha\beta,\text{rel}}$ can be used instead of the absolute model deviation:\cite{Zeng_EnergyFuels_2021_v35_p762}
\begin{align}
    \epsilon_{\bm{F},i,\text{rel}}  (\bm x)
    &= 
    \frac{\lvert \epsilon_{\bm{F},i} (\bm x) \lvert}
    {\lvert \langle \bm F_i (\bm x; \bm \theta_k) \rangle \lvert + \nu}, \\
    \epsilon_{\bm{\Xi},\alpha\beta,\text{rel}}  (\bm x)
    &= 
    \frac{ \epsilon_{\bm{\Xi},\alpha\beta} (\bm x) }
    {\lvert \langle \bm \Xi (\bm x; \bm \theta_k) \rangle \lvert + \nu},
\end{align}
where $\nu$ is a small constant used to protect
an atom where the magnitude of $\bm{F}_i$ or $\bm{\Xi}$ is small from having a large model deviation.

Statistics of $\epsilon_{\bm{F},i}$ and $\epsilon_{\bm{\Xi},{\alpha \beta}}$ can be provided, including the maximum, average, and minimal model deviation:
\begin{align}
    \epsilon_{\bm{F},\text{max}} &= \max_i \epsilon_{\bm{F},i}, \\
    \epsilon_{\bm{F},\text{ave}} &= \frac{1}{N} \sum_i \epsilon_{\bm{F},i}, \\
    \epsilon_{\bm{F},\text{min}} &= \min_i \epsilon_{\bm{F},i}, \\
    \epsilon_{\bm{\Xi},\text{max}} &= \max_{\alpha,\beta} \epsilon_{\bm{\Xi}, \alpha\beta}, \\
    \epsilon_{\bm{\Xi},\text{ave}} &= \frac{1}{9} \sum_{\alpha,\beta=1}^3 \epsilon_{\bm{\Xi}, \alpha\beta}, \\
    \epsilon_{\bm{\Xi},\text{min}} &= \min_{\alpha,\beta} \epsilon_{\bm{\Xi}, \alpha\beta}.
\end{align}
The maximum model deviation of forces $\epsilon_{\bm F,\text{max}}$ in a frame was found to be the best error indicator in a concurrent or active learning algorithm.\cite{Zhang_PhysRevMater_2019_v3_p23804, Zhang_ComputPhysCommun_2020_v253_p107206}

\section{Technical implementation}

In addition to incorporating new powerful features, DeePMD-kit has been designed with the following goals in mind: high performance, high usability, high extensibility, and community engagement.
These goals are crucial for DeePMD-kit to become a widely-used platform across various computational fields.
In this section, we will introduce several technical implementations that have been put in place to achieve these goals.

\subsection{Code architecture}

The DeePMD-kit utilizes TensorFlow's computational graph architecture to construct its DP models\cite{Abadi_2015_tensorflow}, which are composed of various operators implemented with C++, including customized ones such as the environment matrix, Ewald summation, compressed operator, and their backward propagations.
The auto-grad mechanism provided by TensorFlow is used to compute the derivatives of the DP model with respect to the input atomic coordinates and simulation cell tensors.
To optimize performance, some of the critical customized operators are implemented for GPU execution using CUDA or ROCm toolkit libraries.
The DeePMD-kit provides Python, C++, and C APIs for inference, facilitating easy integration with third-party software packages.
As indicated in Figure~\ref{fig:code}, the code of the DeePMD-kit consists of the following modules:

\begin{figure}
    \centering
    \includegraphics[width=\linewidth]{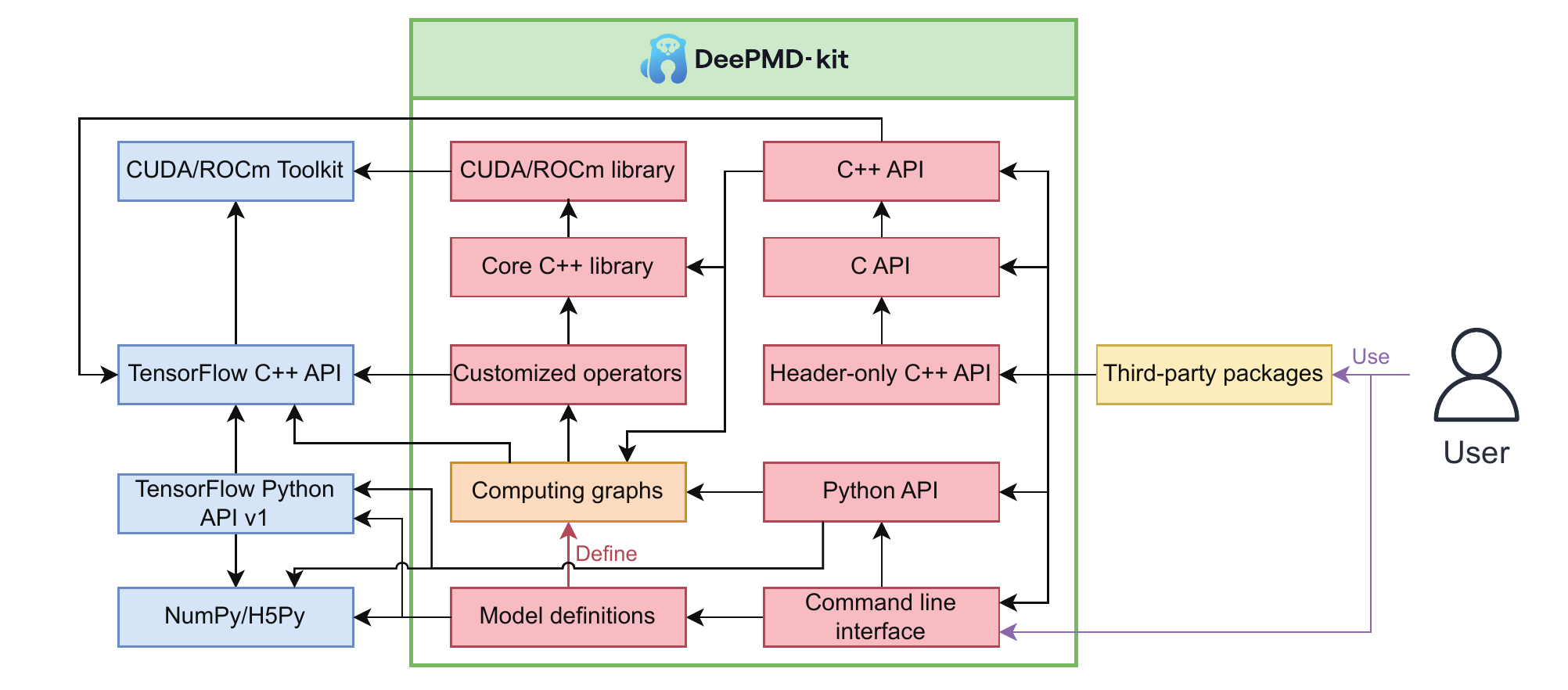}
    
    \caption{The architecture of the DeePMD-kit code. The red boxes are modules within the DeePMD-kit package (the green box), the orange box is computing graphs, the blue boxes are dependencies of the DeePMD-kit, and the yellow box is packages integrated with DeePMD-kit. The direction of the black arrow ${\rm A}\rightarrow {\rm B}$ indicates that module A is dependent on module B. The red and purple arrows represent ``define'' and ``use'', respectively.}
    \label{fig:code}
\end{figure}

\begin{itemize}
\item The core C++ library provides the implementation of customized operators such as the atomic environmental matrix, neighbor lists, and compressed neural networks. It is important to note that the core C++ library is independently built and tested without TensorFlow's C++ interface. 
\item The GPU library (CUDA\cite{Nickolls_2008_CUDA} or ROCm\cite{rocm}), an optional part of the core C++ library, is used to compute customized operators on GPU devices other than CPUs. This library depends on the GPU toolkit library (NVIDIA CUDA Toolkit or AMD ROCm Toolkit) and is also independently built and tested.
\item The DP operators library contains several customized operators not supported by TensorFlow\cite{Abadi_2015_tensorflow}. TensorFlow provides both Python and C++ interfaces to implement some customized operators, with the TensorFlow C++ library packaged inside its Python package.
\item The ``model definitions'' module, written in Python, is used to generate computing graphs composed of TensorFlow operators, DP customized operators, and model parameters organized as ``variables''. The graph can be saved into a file that can be restored for inference. It depends on the TensorFlow Python API (version 1, \texttt{tf.compat.v1}) and other Python dependencies like the NumPy\cite{Harris_Nat_2020_v585_p7825} and H5Py\cite{Collette_2013_Pyh5} packages. 
\item The Python application programming interface (API) is used for inference and can read computing graphs from a file and use the TensorFlow Python API to execute the graph.
\item The C++ API, built upon the TensorFlow C++ interface, does the same thing as the Python API for inference.
\item The C API is a wrapper of the C++ API and provides the same features as the C++ API. Compared to the C++ API, the C API has a more stable application binary interface (ABI) and ensures backward compatibility.
\item The header-only C++ API is a wrapper of the C API and provides the same interface as the C++ API. It has the same stable ABI as the C API but still takes advantage of the flexibility of C++.
\item The command line interface (CLI) is provided to both general users and developers and is used for both training and inference. It depends on the model definition module and the Python API.
\end{itemize}

The CMake build system\cite{Martin_2015_CMake} manages all modules, and the pip and scikit-build\cite{scikit-build} packages are used to distribute DeePMD-kit as a Python package.
Standard Python unit testing framework\cite{Rossum_2018_Python} is used for unit tests on all Python codes, while GoogleTest software\cite{googletest} is used for tests on all C++ codes.
GitHub Actions automates build, test, and deployment pipelines.

\subsection{Performance}

\subsubsection{Hardware acceleration}

In the TensorFlow framework, a static graph combines multiple operators with inputs and outputs. Two kinds of operators are time-consuming during training or inference. The first one is TensorFlow's native operators for neural networks (see Section~\ref{section:NN}) and matrix operations, which have been fully optimized by the TensorFlow framework itself\cite{Abadi_2015_tensorflow}  for both CPU and GPU architectures.
Second, the DeePMD-kit's customized operators 
for computing the \emph{atomic environment} (Eq.~\eqref{eq:rij_loc} and \eqref{eq:rij}) and for the \emph{tabulated inference of the embedding matrix} (Eq.~\eqref{eq:compress}). 
These operators are not supported by the TensorFlow framework
but can be accelerated using OpenMP\cite{openmp}, CUDA\cite{Nickolls_2008_CUDA}, and ROCm\cite{rocm} for parallelization under both CPUs and GPUs.

The operator of the environment matrix includes two steps\cite{Lu_CompPhysCommun_2021_v259_p107624}: formatting the neighbor list and computing the matrix elements of $\mathcal{R}$.
In the formatting step, the neighbors of the atom $i$ are sorted according to their type $\alpha_j$, their distance $r_{ij}$ to atom $i$, and finally their index $j$.
To improve sorting performance on GPUs, the atomic type, distance, and index
are compressed into a 64-bit integer $S \in \mathbb{N}$ used for sorting:
\begin{equation}
    S = \alpha_j \times {10}^{15} + \lfloor r_{ij} \times {10}^8 \rfloor \times {10}^5 + j.
\end{equation}
The sorted neighbor index is decompressed from the sorted $S$ and then used to format the neighbor list.

\subsubsection{MPI implementation for multi-device training and MD simulations}
Users may prefer to utilize multiple CPU cores, GPUs, or hardware across multiple nodes to achieve faster performance and larger memory during training or molecular dynamics (MD) simulations.
To facilitate this, DeePMD-kit has added message-passing interface (MPI) implementation\cite{openmpi,mpich2} for multi-device training and MD simulations in two ways, which are described below.

Multi-device training is conducted with the help of Horovod, a distributed training framework\cite{Sergeev_2018_horovod}. 
Horovod works in the data-parallel mode by equally distributing a batch of data among workers along the axis of the batch size $\mathcal{B}$.\cite{Goyal_arXiv_2017}
During training, each worker consumes sliced input records at different offsets, and only the trainable parameter gradients are averaged with peers. This design avoids batch size and tensor shape conflicts and reduces the number of bytes that need to be communicated among processes.
The mpi4py package\cite{Dalcin_2021_ComputSciEng_v23_p47} is used to remove redundant logs.

Multi-device MD simulations are implemented by utilizing the existing parallelism features of third-party MD packages.
For example, LAMMPS enables parallelism across CPUs by optimizing partitioning, communication, and neighbor lists.\cite{Thompson_ComputPhysCommun_2022_v271_p108171}
AMBER builds a similar neighbor list in the interface to DeePMD-kit.\cite{Case_amber_2020,Zeng_JChemTheoryComput_2021_v17_p6993,Liang_BookCharp_MultiscaleModeling_2023_p6-1}
DeePMD-kit supports local atomic environment calculation and accepts the neighbor list $n(i)$ from other software to replace the native neighbor list calculation.\cite{Lu_CompPhysCommun_2021_v259_p107624}
In a device, the neighbors from other devices are considered ``ghost'' atoms that do not contribute atomic energy $E_i$ to this device's total potential energy $E$.

\subsubsection{Non-von Neumann molecular dynamics (NVNMD)}

When performing molecular dynamics (MD) simulations on CPUs and GPUs, a large majority of time and energy (e.g., more than 95\%) is consumed by the DP model inference. This inference process is limited by the ``memory wall'' and ``power wall'' bottlenecks of von Neumann (vN) architecture, which means that a significant amount of time and energy (e.g., over 90\%) is wasted on data transfer between the processor and memory. As a result, it is difficult to improve computational efficiency.

To address these challenges, non-von Neumann molecular dynamics (NVNMD) uses a non-von Neumann (NvN) architecture chip to accelerate inference. The NvN chip contains processing and memory units that can be used to implement the DP algorithm. In the NvN chip, the hardware algorithm runs fully pipelined. The model parameters are stored in on-chip memory after being loaded from off-chip memory during the initialization process. Therefore, two components of data shuttling are avoided: (1) reading/writing the intermediate results from/to off-chip memory and (2) loading model parameters from off-chip memory during the calculation process. As a result, the DP model ensures high accuracy with NVNMD, while the NvN chip ensures high computational efficiency. For more details, see Ref.~\citenum{Mo_npjComputMater_2022_v8_p107}.

\subsection{Usability}
DeePMD-kit's features and arguments have grown rapidly with more and more development.
To address this issue, we have introduced Sphinx\cite{sphinx} and Doxygen\cite{doxygen} to manage and generate documentation for developers from docstrings in the code.
We use the DArgs package (see Section~\ref{sec:dargs}) to automatically generate Sphinx documentation for user input arguments.
The documentation is currently hosted on Read the Docs (\url{https://docs.deepmodeling.org/projects/deepmd/}).
Furthermore, we strive to make the error messages raised by DeePMD-kit clear to users. In addition, the GitHub Discussion forum allows users to ask questions and receive answers.
Recently, several tutorials have been published\cite{Liang_BookCharp_MultiscaleModeling_2023_p6-1,Zeng_BookChap_QuantChemML_2022_p279} to help new users quickly learn DeePMD-kit.

\subsubsection{Easy installation}
As shown in Figure~\ref{fig:code}, DeePMD-kit has dependencies on both Python and C++ libraries of TensorFlow, which can make it difficult and time-consuming for new users to build TensorFlow and DeePMD-kit from the source code.
Therefore, we provide compiled binary packages that are distributed via pip, Conda (DeepModeling and conda-forge\cite{conda_forge_community_2015_4774217} channels), Docker, and offline packages for Linux, macOS, and Windows platforms.
With the help of these pre-compiled binary packages, users can install DeePMD-kit in just a few minutes.
These binary packages include DeePMD-kit's LAMMPS plugin, i-PI driver, and GROMACS patch.
As LAMMPS provides a plugin mode in its latest version, DeePMD-kit's LAMMPS plugin can be compiled without having to re-compile LAMMPS.\cite{Thompson_ComputPhysCommun_2022_v271_p108171}
We offer a compiled binary package that includes the C API and the header-only C++ API, making it simpler to integrate with sophisticated software like AMBER.\cite{Case_amber_2020,Zeng_JChemTheoryComput_2021_v17_p6993,Liang_BookCharp_MultiscaleModeling_2023_p6-1}

\subsubsection{User interface}

DeePMD-kit offers a command line interface (CLI) for training, freezing, and testing models. In addition to CLI arguments, users must provide a JSON\cite{Pezoa_2016_JSON} or YAML\cite{Sinha_2000_YAML} file with completed arguments for components listed in Section~\ref{sec:theory}. The DArgs package (see Section~\ref{sec:dargs}) parses these arguments to check if user input is correct. An example of how to use the user interface is provided in Ref.~\citenum{Liang_BookCharp_MultiscaleModeling_2023_p6-1}.
Users can also use DP-GUI (see Section~\ref{sec:dpgui}) to fill in arguments in an interactive web page and save them to a JSON\cite{Pezoa_2016_JSON} file.

DeePMD-kit provides an automatic algorithm that assists new users in deciding on several arguments.
For example, the automatic batch size $\mathcal{B}$ determines the maximum batch size during training or inferring to fully utilize memory on a GPU card.
The automatic neighbor size $N_c$ determines the maximum number of neighbors by stating the training data to reduce model memory usage.
The automatic probability determines the probability of using a system during training.
These automatic arguments reduce the difficulty of learning and using the DeePMD-kit.

\subsubsection{Input data}
To train and test models, users are required to provide fitting data in a specified format.
DeePMD-kit supports two file formats for data input: NumPy binary files\cite{Harris_Nat_2020_v585_p7825} and HDF5 files\cite{hdf5}.
These formats are designed to offer superior performance when read by the program with parallel algorithms compared to text files.
HDF5 files have the advantage of being able to store multiple arrays in a single file, making them easier to transfer between machines. The Python package ``DP-Data'' (see Section~\ref{sec:dpdata}) can generate these files from the output of an electronic calculation package.

\subsubsection{Model visualization}
DeePMD-kit supports most of the visualization features offered by TensorBoard\cite{Abadi_2015_tensorflow}, such as tracking and visualizing metrics, viewing the model graph, histograms of tensors, summaries of trainable variables, and debugging profiles.

\subsection{Extensibility}

\subsubsection{Application programming interface and third-party software}

DeePMD-kit offers various APIs, including the Python, C++, C, and header-only C++ API, as well as a command-line interface (CLI), as shown in Figure~\ref{fig:code}.
These APIs are primarily used for inference by developers and high-level users in different situations.
Sphinx\cite{sphinx} generates the API details in the documentation.

These APIs can be easily accessed by various third-party software.
The Python API, for instance, is utilized by third-party Python packages, such as  ASE\cite{Larsen_JPhysCondensMat_2017_v29_p273002} and dpdata (see Section~\ref{sec:dpdata}).
The C++, C, or header-only C++ API has also been integrated into several third-party MD packages, such as
LAMMPS\cite{Plimpton_JCompPhys_1995_v117_p1,Thompson_ComputPhysCommun_2022_v271_p108171},
i-PI\cite{Kapil_CompPhysCommun_2019_v236_p214},
GROMACS\cite{Abraham_SoftX_2015_v1_p19},
AMBER\cite{Case_amber_2020,Zeng_JChemTheoryComput_2021_v17_p6993,Liang_BookCharp_MultiscaleModeling_2023_p6-1},
OpenMM\cite{Eastman_PLoSCompBio_2017_v13_p7,openmm_deepmd_plugin},
and ABACUS\cite{Li_ComputationalMaterialsScience_2016_v112_p503}.
Moreover, the CLI is called by various third-party workflow packages, such as
DP-GEN\cite{Zhang_ComputPhysCommun_2020_v253_p107206}
and MLatom\cite{Dral_TopCurrChem(Cham)_2021_v379_p27}.
While the ASE calculator, the LAMMPS plugin, the i-PI driver, and the GROMACS patch are developed within the DeePMD-kit code, others are distributed separately. 
By integrating these APIs into their programs, researchers can perform simulations and minimization, without being restricted by DeePMD-kit's software features.\cite{CalegariAndrade_ChemSci_2020_v11_p2335,Piaggi_ProcNatlAcadSciUSA_2022_v119_pe2207294119,Giese_JChemTheoryComput_2022_v18_p4304,Zeng_JChemTheoryComput_2023_v19_p1261}
Additionally, they can combine DP models with other potentials outside the DeePMD-kit package if necessary.\cite{Zeng_JChemTheoryComput_2021_v17_p6993,Achar_JChemTheoryComput_2022_v18_p3593,Zeng_JChemTheoryComput_2023_v19_p1261}

\subsubsection{Customized plugins}

DeePMD-kit is built with an object-oriented design, and each component discussed in Section \ref{sec:theory} corresponds to a Python class.
One of the advantages of this design is the availability of a plugin system for these components.
With this plugin system, developers can create and incorporate their customized components, without having to modify the DeePMD-kit package. This approach expedites the realization of their ideas.
Moreover, the plugin system facilitates the addition of new components within the DeePMD-kit package itself.

\subsection{DeepModeling Community}
\label{sec:dpti}\label{sec:dpdata}\label{sec:dpdispatcher}\label{sec:dargs}\label{sec:dpgui}
DeePMD-kit is a free and open-source software licensed under the LGPL-3.0 license, enabling developers to modify and incorporate DeePMD-kit into their own packages.
Serving as the core, DeePMD-kit led to the formation of an open-source community named DeepModeling in 2021, which manages open-source packages for scientific computing.
Since then, numerous open-source packages for scientific computing have either been created or joined the DeepModeling community, such as
DP-GEN\cite{Zhang_ComputPhysCommun_2020_v253_p107206},
DeePKS-kit\cite{Chen_ComputerPhysicsCommunications_2023_v282_p108520},
DMFF\cite{Wang_2022_DMFF},
ABACUS\cite{Li_ComputationalMaterialsScience_2016_v112_p503},
DeePH \cite{Li_NatComputSci_2022_v2_p367}, 
and DeepFlame\cite{Chen_2023_DeepFlame},
among others, whether directly or indirectly related to DeePMD-kit.
The DeepModeling packages that are related to DeePMD-kit are listed below.
\begin{enumerate}
\item Deep Potential GENerator (DP-GEN)\cite{Zhang_ComputPhysCommun_2020_v253_p107206} is a package that implements the concurrent learning procedure\cite{Zhang_PhysRevMater_2019_v3_p23804} and is capable of generating uniformly accurate DP models with minimal human intervention and computational cost. DP-GEN2 is the next generation of this package, built on the workflow platform Dflow.
\item
Deep Potential Thermodynamic Integration (DP-Ti) is a Python package that enables users to calculate free energy, perform thermodynamic integration, and determine pressure-temperature phase diagrams for materials with DP models.
\item
DP-Data is a Python package that helps users convert atomistic data between different formats and calculate atomistic data through electronic calculation and MLP packages. It can be used to generate training data files for DeePMD-kit and visualize structures via 3Dmol.js\cite{Rego_Bioinformatics_2015_v31_p1322}. The package supports a plugin system and is compatible with ASE\cite{Larsen_JPhysCondensMat_2017_v29_p273002}, allowing it to support any data format without being limited by the package's code.
\item
DP-Dispatcher is a Python package used to generate input scripts for high-performance computing (HPC) schedulers, submit them to HPC systems, and monitor their progress until completion. It was originally developed as part of the DP-GEN package\cite{Zhang_ComputPhysCommun_2020_v253_p107206}, but has since become an independent package that serves other packages.
\item
DArgs is a Python package that manages and filters user input arguments. It provides a Sphinx\cite{sphinx} extension to generate documentation for arguments.
\item
DP-GUI is a web-based graphical user interface (GUI) built with the Vue.js framework\cite{vuejs} It allows users to fill in arguments interactively on a web page and save them to a JSON\cite{Pezoa_2016_JSON} file. DArgs is used to provide details and documentation of arguments in the GUI.
\end{enumerate}

\section{Benchmarking}

We performed benchmarking on various potential energy models with different descriptors on multiple datasets to showcase the precision and performance of descriptors developed within the DeePMD-kit package. 
The datasets we used included water\cite{Zhang_PhysRevLett_2018_v120_p143001,Ko_MolecularPhysics_2019_v117_p3269}, copper (Cu)\cite{Zhang_ComputPhysCommun_2020_v253_p107206}, high entropy alloys (HEA)\cite{jiang2023hybrid}, OC2M subset in Open Catalyst 2020 (OC20)\cite{Chanussot_ACSCatal_2021_v11_p6059,gasteiger2022gemnetoc}, Small-Molecule/Protein Interaction Chemical Energies (SPICE)\cite{Eastman_SciData_2023_v10_p11}, and dipeptides subset in SPICE\cite{Eastman_SciData_2023_v10_p11}.
We split all the datasets into a training set containing 95\% of the data and a validation set containing the remaining 5\% of the data.

We compared various descriptors, including the local frame (\texttt{loc\textunderscore frame}),
two-body embedding full-information DeepPot-SE (\texttt{se\textunderscore e2\textunderscore a}), a hybrid descriptor with two-body embedding full- and radial-information DeepPot-SE (\texttt{se\textunderscore e2\textunderscore a+se\textunderscore e2\textunderscore r}), a hybrid descriptor with two-body embedding full-information and three-body embedding DeepPot-SE (\texttt{se\textunderscore e2\textunderscore a+se\textunderscore e3}), and an attention-based descriptor (\texttt{se\textunderscore atten}).
In all models, we set $r_s$ to 0.5~\AA{}, $M_<$ to 16, and $L_a$ to 2, if applicable.
We used (25,50,100) neurons for two-body embedding networks $\mathcal{N}_{e,2}$, (2,4,8) neurons for three-body embedding networks $\mathcal{N}_{e,3}$, and (240,240,240,1) neurons for fitting networks $\mathcal{F}_0$.
In the full-information part (\texttt{se\textunderscore e2\textunderscore a})
of the hybrid descriptor with two-body embedding full-information and radius-information DeepPot-SE (\texttt{se\textunderscore e2\textunderscore a+se\textunderscore e2\textunderscore r}) and the two-body embedding part (\texttt{se\textunderscore e2\textunderscore a}) of the hybrid descriptor with two-body full-information and three-body DeepPot-SE (\texttt{se\textunderscore e2\textunderscore a+se\textunderscore e3}), we set $r_c$ to 4~\AA{}. For the OC2M system, we set $r_c$ to 9~\AA{}, while under other situations, we set $r_c$ to 6~\AA{}.
We trained each model for a fixed number of steps (\num{1000000} for water, Cu, and dipeptides, \num{16000000} for HEA, and \num{10000000} for OC2M and SPICE) using neural networks in double floating precision (FP64) and single floating precision (FP32) separately.
We used the LAMMPS package\cite{Thompson_ComputPhysCommun_2022_v271_p108171} to perform MD simulations for water, Cu, and HEA with as many atoms as possible.
We compared the performance of compressed models with that of the original model where applicable.\cite{Lu_CompPhysCommun_2021_v259_p107624}.
The platforms used to benchmark performance included 128-core AMD EPYC 7742, NVIDIA GeForce RTX 3080 Ti, NVIDIA Tesla V100, NVIDIA Tesla A100, AMD Instinct MI250, and Xilinx Virtex Ultrascale+ VU9P FPGA for NVNMD only\cite{Mo_npjComputMater_2022_v8_p107}.
We note that currently, the model compression feature only supports \texttt{se\textunderscore e2\textunderscore a}, \texttt{se\textunderscore e2\textunderscore r}, and \texttt{se\textunderscore e3} descriptors, and NVNMD only supports regular \texttt{se\textunderscore e2\textunderscore a} for systems with no more than 4 chemical species in FP64 precision.

\begin{table*}
  \small
  \caption[]{Mean absolute errors (MAE) and root mean square errors (RMSE) in the energy per atom (E, meV/atom) and forces ({\bf F}, meV/\AA) for water, Cu, HEA, OC2M, dipeptides, and SPICE validation sets. The underline donates the best model in an indicator.
  }
  \begin{ruledtabular}
    \begin{tabular}{ll|rr|rr|rr|rr|rr}
    \multirow{2}{*}{System} & \multirow{2}{*}{Indicator} & \multicolumn{2}{c|}{\texttt{loc\textunderscore frame}} & \multicolumn{2}{c|}{\texttt{se\textunderscore e2\textunderscore a}} & \multicolumn{2}{c|}{\texttt{se\textunderscore e2\textunderscore a+se\textunderscore e2\textunderscore r}} & \multicolumn{2}{c|}{\texttt{se\textunderscore e2\textunderscore a+se\textunderscore e3}} & \multicolumn{2}{c}{\texttt{se\textunderscore atten}} \\
    & & \multicolumn{1}{c}{FP64} & \multicolumn{1}{c|}{FP32} & \multicolumn{1}{c}{FP64} & \multicolumn{1}{c|}{FP32} & \multicolumn{1}{c}{FP64} & \multicolumn{1}{c|}{FP32} & \multicolumn{1}{c}{FP64} & \multicolumn{1}{c|}{FP32} & \multicolumn{1}{c}{FP64} & \multicolumn{1}{c}{FP32} \\
      \hline
    \multirow{4}{*}{Water}
     & E MAE        & \float{5.50E-04} & \underline{\float{5.41E-04}} & \float{7.79E-04} & \float{7.67E-04} & \float{7.27E-04} & \float{7.60E-04} & \float{7.46E-04} & \float{7.36E-04} & \float{1.26E-03} & \float{9.72E-04} \\
     & E RMSE       & \float{6.93E-04} & \underline{\float{6.89E-04}} & \float{9.93E-04} & \float{9.76E-04} & \float{9.28E-04} & \float{9.70E-04} & \float{9.87E-04} & \float{9.72E-04} & \float{1.50E-03} & \float{1.18E-03} \\
     & {\bf F} MAE  & \float{2.98E-02} & \underline{\float{2.93E-02}} & \float{3.66E-02} & \float{3.62E-02} & \float{3.65E-02} & \float{3.75E-02} & \float{3.44E-02} & \float{3.40E-02} & \float{3.24E-02} & \float{3.10E-02} \\
     & {\bf F} RMSE & \float{4.00E-02} & \underline{\float{3.92E-02}} & \float{4.90E-02} & \float{4.84E-02} & \float{4.86E-02} & \float{5.00E-02} & \float{4.65E-02} & \float{4.59E-02} & \float{4.44E-02} & \float{4.23E-02} \\
      \hline
    \multirow{4}{*}{Cu}
     & E MAE        & \float{6.32E-03} & \float{8.46E-03} & \float{2.18E-03} & \float{2.06E-03} & \float{3.58E-03} & \float{3.65E-03} & \underline{\float{1.84E-03}} & \float{1.86E-03} & \float{2.41E-03} & \float{2.71E-03} \\
     & E RMSE       & \float{1.27E-02} & \float{1.92E-02} & \float{2.95E-03} & \float{2.82E-03} & \float{4.83E-03} & \float{4.93E-03} & \underline{\float{2.54E-03}} & \float{2.58E-03} & \float{3.24E-03} & \float{3.60E-03} \\
     & {\bf F} MAE  & \float{4.89E-02} & \float{5.53E-02} & \float{1.02E-02} & \float{1.03E-02} & \float{1.20E-02} & \float{1.23E-02} & \float{9.76E-03} & \underline{\float{9.63E-03}} & \float{9.65E-03} & \float{9.70E-03} \\
     & {\bf F} RMSE & \float{8.47E-02} & \float{1.05E-01} & \float{1.77E-02} & \float{1.79E-02} & \float{2.14E-02} & \float{2.20E-02} & \float{1.68E-02} & \underline{\float{1.66E-02}} & \float{1.69E-02} & \float{1.69E-02} \\
      \hline
    \multirow{4}{*}{HEA}
     & E MAE        & \dots & \dots & \float{1.09E-02} & \float{1.08E-02} & \float{8.15E-03} & \float{8.77E-03} & \float{8.22E-03} & \float{1.14E-02} & \underline{\float{3.09E-03}} & \float{3.52E-03} \\
     & E RMSE       & \dots & \dots & \float{1.54E-02} & \float{1.53E-02} & \float{1.35E-02} & \float{1.45E-02} & \float{1.21E-02} & \float{1.72E-02} & \underline{\float{5.48E-03}} & \float{6.44E-03} \\
     & {\bf F} MAE  & \dots & \dots & \float{9.37E-02} & \float{9.39E-02} & \float{1.16E-01} & \float{1.13E-01} & \float{9.33E-02} & \float{1.22E-01} & \underline{\float{5.55E-02}} & \float{5.68E-02} \\
     & {\bf F} RMSE & \dots & \dots & \float{1.34E-01} & \float{1.37E-01} & \float{1.63E-01} & \float{1.58E-01} & \float{1.36E-01} & \float{1.80E-01} & \underline{\float{9.07E-02}} & \float{9.83E-02} \\
      \hline
    \multirow{4}{*}{OC2M}
     & E MAE        & \dots & \dots & \dots & \dots & \dots & \dots & \dots & \dots & \float{1.10E-02} & \underline{\float{1.04E-02}}\\ 
     & E RMSE       & \dots & \dots & \dots & \dots & \dots & \dots & \dots & \dots & \float{1.51E-02} & \underline{\float{1.43E-02}} \\
     & {\bf F} MAE  & \dots & \dots & \dots & \dots & \dots & \dots & \dots & \dots & \float{9.86E-02} & \underline{\float{9.56E-02}} \\
     & {\bf F} RMSE & \dots & \dots & \dots & \dots & \dots & \dots & \dots & \dots & \float{1.55E-01} & \underline{\float{1.48E-01}} \\
      \hline
    \multirow{4}{*}{Dipeptides}
     & E MAE        & \dots & \dots & \underline{\float{5.47E-03}} & \float{5.52E-03} & \float{6.18E-03} & \float{6.11E-03} & \float{7.57E-03} & \float{9.15E-03} & \float{7.05E-03} & \float{6.90E-03} \\
     & E RMSE       & \dots & \dots & \underline{\float{9.51E-03}} & \float{9.60E-03} & \float{1.25E-02} & \float{1.17E-02} & \float{1.49E-02} & \float{1.68E-02} & \float{1.28E-02} & \float{1.27E-02} \\
     & {\bf F} MAE  & \dots & \dots & \float{6.78E-02} & \float{6.78E-02} & \float{6.83E-02} & \float{7.00E-02} & \float{1.10E-01} & \float{1.38E-01} & \float{6.94E-02} & \underline{\float{6.76E-02}} \\
     & {\bf F} RMSE & \dots & \dots & \float{9.79E-02} & \float{9.77E-02} & \float{9.86E-02} & \float{1.01E-01} & \float{1.60E-01} & \float{2.22E-01} & \float{9.97E-02} & \underline{\float{9.67E-02}} \\
      \hline
    \multirow{4}{*}{SPICE}
     & E MAE        & \dots & \dots & \dots & \dots & \dots & \dots & \dots & \dots & \underline{\float{1.53E-02}} & \float{1.53E-02} \\
     & E RMSE       & \dots & \dots & \dots & \dots & \dots & \dots & \dots & \dots & \float{8.09E-02} & \underline{\float{7.83E-02}} \\
     & {\bf F} MAE  & \dots & \dots & \dots & \dots & \dots & \dots & \dots & \dots & \underline{\float{1.10E-01}} & \float{1.12E-01} \\
     & {\bf F} RMSE & \dots & \dots & \dots & \dots & \dots & \dots & \dots & \dots & \underline{\float{2.33E-01}} & \float{2.34E-01} \\
    \end{tabular}
\end{ruledtabular}
    \justifying
    \noindent
    \label{t:error}\vspace{-0.5cm}
\end{table*}

\begin{table*}
  \small
  \centering
  \caption[]{Training performance (ms/step) for water, Cu, HEA, OC2M, dipeptides, and SPICE systems. ``FP64'' means double floating precision, ``FP32'' means single floating precision, and ``FP64c'' and ``FP32c'' mean the compressed training\cite{Lu_JChemTheoryComput_2022_v18_p5559} for double and single floating precision, respectively. ``EPYC'' performed on 128 AMD EPYC 7742 cores, ``3080 Ti'' performed on an NVIDIA GeForce RTX 3080 Ti card, ``V100'' performed on an NVIDIA Tesla V100 card, ``A100'' performed on an NVIDIA Tesla A100 card, and ``MI250'' performed on an AMD Instinct MI250 Graphics Compute Die (GCD).
  }
  \resizebox*{\textwidth}{!}{%
    \begin{tabular}{l|ll|rr|rrrr|rrrr|rrrr|rr}
    \hline \hline
    \multirow{2}{*}{System} & & \multirow{2}{*}{Hardware} & \multicolumn{2}{c|}{\texttt{loc\textunderscore frame}} & \multicolumn{4}{c|}{\texttt{se\textunderscore e2\textunderscore a}} & \multicolumn{4}{c|}{\texttt{se\textunderscore e2\textunderscore a+se\textunderscore e2\textunderscore r}} & \multicolumn{4}{c|}{\texttt{se\textunderscore e2\textunderscore a+se\textunderscore e3}} & \multicolumn{2}{c}{\texttt{se\textunderscore atten}} \\
    & & & \multicolumn{1}{c}{FP64} & \multicolumn{1}{c|}{FP32} & \multicolumn{1}{c}{FP64} & \multicolumn{1}{c}{FP32} & \multicolumn{1}{c}{FP64c} & \multicolumn{1}{c|}{FP32c} & \multicolumn{1}{c}{FP64} & \multicolumn{1}{c}{FP32} & \multicolumn{1}{c}{FP64c} & \multicolumn{1}{c|}{FP32c} & \multicolumn{1}{c}{FP64} & \multicolumn{1}{c}{FP32} & \multicolumn{1}{c}{FP64c} & \multicolumn{1}{c|}{FP32c} & \multicolumn{1}{c}{FP64} & \multicolumn{1}{c}{FP32} \\
      \hline
    \multirow{5}{*}{Water}
     & 
       & EPYC & \float{1.47E-02} & \float{9.20E-03} & \float{9.73E-02} & \float{4.50E-02} & \float{2.84E-02} & \float{1.62E-02} & \float{6.37E-02} & \float{3.25E-02} & \float{2.99E-02} & \float{1.54E-02} & \float{1.41E-01} & \float{8.52E-02} & \float{3.40E-02} & \float{2.06E-02} & \float{1.2132} & \float{0.3829} \\
     & & 3080 Ti & \float{7.00E-03} & \float{4.80E-03} & \float{2.46E-02} & \float{1.03E-02} & \float{9.70E-03} & \float{6.40E-03} & \float{2.63E-02} & \float{1.16E-02} & \float{1.20E-02}  & \float{8.20E-03} & \float{5.28E-02} & \float{1.72E-02} & \float{1.63E-02} & \float{6.80E-03} & \float{1.99E-01} & \float{2.69E-02} \\
     & & V100 & \float{7.90E-03} & \float{8.50E-03} & \float{1.11E-02} & \float{8.20E-03} & \float{5.90E-03} & \float{4.80E-03} & \float{1.36E-02} & \float{1.09E-02} & \float{6.90E-03} & \float{6.40E-03} & \float{2.35E-02} & \float{1.40E-02} & \float{8.60E-03} & \float{7.30E-03} & \float{6.96E-02} & \float{3.17E-02} \\
     & & A100 & \float{1.07E-02} & \float{1.00E-02} & \float{8.20E-03} & \float{9.30E-03} & \float{4.90E-03} & \float{5.70E-03} & \float{1.45E-02} & \float{1.08E-02} & \float{7.80E-03} & \float{6.30E-03} & \float{2.45E-02} & \float{1.20E-02} & \float{7.50E-03} & \float{7.20E-03} & \float{3.08E-02} & \float{2.12E-02} \\
     & & MI250 & \float{1.17E-02} & \float{1.09E-02} & \float{2.03E-02} & \float{1.31E-02} & \float{7.70E-03} & \float{7.00E-03} & \float{2.73E-02} & \float{1.97E-02} & \float{1.15E-02} & \float{1.09E-02} & \float{2.78E-01} & \float{2.77E-02} & \float{1.28E-02} & \float{1.12E-02} & \float{1.25E-01} & \float{3.17E-02} \\
     \cline{2-19}
      \hline
    \multirow{5}{*}{Cu}
     & 
       & EPYC & \float{4.90E-03} & \float{3.30E-03} & \float{3.37E-02} & \float{1.28E-02} & \float{8.00E-03} & \float{5.40E-03} & \float{1.99E-02} & \float{1.00E-02} & \float{1.05E-02} & \float{5.30E-03} & \float{4.55E-02} & \float{2.42E-02} & \float{9.10E-03} & \float{6.50E-03} & \float{2.26E-01} & \float{8.91E-02} \\
     & & 3080 Ti & \float{3.20E-03} & \float{2.20E-03} & \float{6.50E-03} & \float{5.10E-03} & \float{4.60E-03} & \float{3.90E-03} & \float{8.70E-03} & \float{6.30E-03} & \float{5.90E-03} & \float{3.40E-03} & \float{1.18E-02} & \float{4.80E-03} & \float{7.20E-03} & \float{5.70E-03} & \float{3.68E-02} & \float{8.80E-03} \\
     & & V100 & \float{3.20E-03} & \float{3.80E-03} & \float{4.20E-03} & \float{4.80E-03} & \float{3.20E-03} & \float{3.70E-03} & \float{6.50E-03} & \float{5.30E-03} & \float{5.50E-03} & \float{4.10E-03} & \float{7.90E-03} & \float{5.60E-03} & \float{6.00E-03} & \float{5.80E-03} & \float{1.56E-02} & \float{1.19E-02} \\
     & & A100 & \float{4.00E-03} & \float{3.90E-03} & \float{3.80E-03} & \float{3.70E-03} & \float{3.10E-03} & \float{3.00E-03} & \float{5.40E-03} & \float{5.30E-03} & \float{4.10E-03} & \float{4.10E-03} & \float{8.00E-03} & \float{5.60E-03} & \float{4.80E-03} & \float{4.60E-03} & \float{1.16E-02} & \float{1.12E-02} \\
     & & MI250 & \float{4.80E-03} & \float{4.90E-03} & \float{6.90E-03} & \float{6.40E-03} & \float{5.10E-03} & \float{5.00E-03} & \float{9.10E-03} & \float{9.40E-03} & \float{7.40E-03} & \float{7.00E-03} & \float{4.99E-02} & \float{1.01E-02} & \float{8.00E-03} & \float{7.30E-03} & \float{2.36E-02} & \float{1.86E-02} \\
     \cline{2-19}
      \hline
    \multirow{5}{*}{HEA}
     & 
       & EPYC & \dots & \dots & \float{5.34E-02} & \float{3.05E-02} & \float{1.94E-02} & \float{1.22E-02} & \float{5.23E-02} & \float{2.93E-02} & \float{2.77E-02} & \float{1.67E-02} & \float{8.37E-02} & \float{5.11E-02} & \float{2.66E-02} & \float{1.57E-02} & \float{1.59E-01} & \float{6.01E-02} \\
     & & 3080 Ti & \dots & \dots & \float{3.84E-02} & \float{2.52E-02} & \float{1.12E-02} & \float{9.10E-03} & \float{7.14E-02} & \float{4.18E-02} & \float{1.63E-02} & \float{1.27E-02} & \float{9.36E-02} & \float{4.10E-02} & \float{1.97E-02} & \float{1.50E-02} & \float{3.59E-02} & \float{9.10E-03} \\
     & & V100 & \dots & \dots & \float{3.32E-02} & \float{2.98E-02} & \float{1.18E-02} & \float{1.11E-02} & \float{6.32E-02} & \float{4.74E-02} & \float{1.75E-02} & \float{1.65E-02} & \float{6.55E-02} & \float{4.96E-02} & \float{2.74E-02} & \float{1.87E-02} & \float{1.56E-02} & \float{1.19E-02} \\
     & & A100 & \dots & \dots & \float{3.05E-02} & \float{2.86E-02} & \float{1.09E-02} & \float{1.04E-02} & \float{5.16E-02} & \float{6.74E-02} & \float{1.69E-02} & \float{2.12E-02} & \float{6.17E-02} & \float{5.29E-02} & \float{1.86E-02} & \float{1.88E-02} & \float{1.17E-02} & \float{1.15E-02} \\
     & & MI250 & \dots & \dots & \float{4.88E-02} & \float{4.27E-02} & \float{1.85E-02} & \float{1.80E-02} & \float{7.23E-02} & \float{6.93E-02} & \float{2.87E-02} & \float{2.73E-02} & \float{1.34E-01} & \float{8.84E-02} & \float{3.27E-02} & \float{3.23E-02} & \float{2.16E-02} & \float{1.95E-02} \\
     \cline{2-19}
      \hline
    \multirow{5}{*}{OC2M}
     & 
       & EPYC & \dots & \dots & \dots & \dots & \dots & \dots & \dots & \dots & \dots & \dots & \dots & \dots & \dots & \dots & \float{2.07E+00} & \float{6.25E-01} \\
     & & 3080 Ti & \dots & \dots & \dots & \dots & \dots & \dots & \dots & \dots & \dots & \dots & \dots & \dots & \dots & \dots & \float{3.52E-01} & \float{4.60E-02} \\
     & & V100 & \dots & \dots & \dots & \dots & \dots & \dots & \dots & \dots & \dots & \dots & \dots & \dots & \dots & \dots & \float{1.20E-01} & \float{5.28E-02} \\
     & & A100 & \dots & \dots & \dots & \dots & \dots & \dots & \dots & \dots & \dots & \dots & \dots & \dots & \dots & \dots & \float{5.14E-02} & \float{3.09E-02} \\
     & & MI250 & \dots & \dots & \dots & \dots & \dots & \dots & \dots & \dots & \dots & \dots & \dots & \dots & \dots & \dots & \float{1.71E-01} & \float{5.57E-02} \\
       \hline
    \multirow{5}{*}{Dipeptides}
     & 
       & EPYC & \dots & \dots & \float{4.97E-02} & \float{3.05E-02} & \float{2.12E-02} & \float{1.94E-02} & \float{5.20E-02} & \float{3.53E-02} & \float{3.01E-02} & \float{2.12E-02} & \float{8.95E-02} & \float{6.11E-02} & \float{3.50E-02} & \float{2.12E-02} & \float{2.14E-01} & \float{9.15E-02} \\
     & & 3080 Ti & \dots & \dots & \float{5.48E-02} & \float{3.95E-02} & \float{1.73E-02} & \float{1.13E-02} & \float{9.00E-02} & \float{6.43E-02} & \float{1.90E-02} & \float{1.53E-02} & \float{1.31E-01} & \float{6.77E-02} & \float{2.54E-02} & \float{1.92E-02} & \float{2.61E-02} & \float{1.20E-02} \\
     & & V100 & \dots & \dots & \float{5.41E-02} & \float{5.26E-02} & \float{1.48E-02} & \float{1.48E-02} & \float{8.80E-02} & \float{8.43E-02} & \float{2.05E-02} & \float{2.17E-02} & \float{9.62E-02} & \float{1.03E-01} & \float{3.01E-02} & \float{3.08E-02} & \float{1.43E-02} & \float{1.06E-02} \\
     & & A100 & \dots & \dots & \float{5.02E-02} & \float{5.08E-02} & \float{1.43E-02} & \float{1.43E-02} & \float{8.90E-02} & \float{7.59E-02} & \float{2.07E-02} & \float{1.99E-02} & \float{9.11E-02} & \float{8.27E-02} & \float{2.66E-02} & \float{2.67E-02} & \float{1.32E-02} & \float{1.11E-02} \\
     & & MI250 & \dots & \dots & \float{6.62E-02} & \float{6.78E-02} & \float{2.31E-02} & \float{2.29E-02} & \float{1.17E-01} & \float{1.12E-01} & \float{3.50E-02} & \float{3.24E-02} & \float{1.55E-01} & \float{1.29E-01} & \float{4.59E-02} & \float{4.49E-02} & \float{1.96E-02} & \float{1.68E-02} \\
       \hline
    \multirow{5}{*}{SPICE}
     & 
       & EPYC & \dots & \dots & \dots & \dots & \dots & \dots & \dots & \dots & \dots & \dots & \dots & \dots & \dots & \dots & \float{2.44E-01} & \float{9.80E-02} \\
     & & 3080 Ti & \dots & \dots & \dots & \dots & \dots & \dots & \dots & \dots & \dots & \dots & \dots & \dots & \dots & \dots & \float{3.54E-02} & \float{1.53E-02} \\
     & & V100 & \dots & \dots & \dots & \dots & \dots & \dots & \dots & \dots & \dots & \dots & \dots & \dots & \dots & \dots & \float{1.73E-02} & \float{1.59E-02} \\
     & & A100 & \dots & \dots & \dots & \dots & \dots & \dots & \dots & \dots & \dots & \dots & \dots & \dots & \dots & \dots & \float{1.19E-02} & \float{1.22E-02} \\
     & & MI250 & \dots & \dots & \dots & \dots & \dots & \dots & \dots & \dots & \dots & \dots & \dots & \dots & \dots & \dots & \float{2.90E-02} & \float{2.41E-02}\\
    \hline \hline
    \end{tabular}%
}
    \justifying
    \noindent
    \label{t:performance-training}\vspace{-0.5cm}
\end{table*}

\begin{table*}
  \small
  \centering
  \caption[]{MD performance ($\mu$s/step/atom) for water, Cu, and HEA systems. ``FP64'' means double floating precision, ``FP32'' means single floating precision, and ``FP64c'' and ``FP32c'' mean the compressed model\cite{Lu_JChemTheoryComput_2022_v18_p5559} for double and single floating precision, respectively. ``EPYC'' performed on 128 AMD EPYC 7742 cores, ``3080 Ti'' performed on an NVIDIA GeForce RTX 3080 Ti card, ``V100'' performed on an NVIDIA Tesla V100 card, ``A100'' performed on an NVIDIA Tesla A100 card, ``MI250'' performed on an AMD Instinct MI250 Graphics Compute Die (GCD), and ``VU9P'' performed NVNMD\cite{Mo_npjComputMater_2022_v8_p107} on a Xilinx Virtex Ultrascale+ VU9P FPGA board.
  }
  \resizebox*{\textwidth}{!}{%
    \begin{tabular}{l|ll|rr|rrrr|rrrr|rrrr|rr}
    \hline \hline
    \multirow{2}{*}{System} & & \multirow{2}{*}{Hardware} & \multicolumn{2}{c|}{\texttt{loc\textunderscore frame}} & \multicolumn{4}{c|}{\texttt{se\textunderscore e2\textunderscore a}} & \multicolumn{4}{c|}{\texttt{se\textunderscore e2\textunderscore a+se\textunderscore e2\textunderscore r}} & \multicolumn{4}{c|}{\texttt{se\textunderscore e2\textunderscore a+se\textunderscore e3}} & \multicolumn{2}{c}{\texttt{se\textunderscore atten}} \\
    & & & \multicolumn{1}{c}{FP64} & \multicolumn{1}{c|}{FP32} & \multicolumn{1}{c}{FP64} & \multicolumn{1}{c}{FP32} & \multicolumn{1}{c}{FP64c} & \multicolumn{1}{c|}{FP32c} & \multicolumn{1}{c}{FP64} & \multicolumn{1}{c}{FP32} & \multicolumn{1}{c}{FP64c} & \multicolumn{1}{c|}{FP32c} & \multicolumn{1}{c}{FP64} & \multicolumn{1}{c}{FP32} & \multicolumn{1}{c}{FP64c} & \multicolumn{1}{c|}{FP32c} & \multicolumn{1}{c}{FP64} & \multicolumn{1}{c}{FP32} \\
      \hline
    \multirow{6}{*}{Water}
     & 
       & EPYC & \floatm{1.25E-06} & \floatm{6.99E-07} & \floatm{1.93E-05} & \floatm{8.73E-06} & \floatm{3.89E-06} & \floatm{2.61E-06} & \floatm{8.33E-06} & \floatm{3.43E-06} & \floatm{3.78E-06} & \floatm{1.86E-06} & \floatm{3.72E-05} & \floatm{1.51E-05} & \floatm{5.04E-06} & \floatm{3.63E-06} & \floatm{2.21E-04} & \floatm{8.38E-05} \\
     & & 3080 Ti & \floatm{1.29E-05} & \floatm{8.63E-06} & \floatm{2.90E-05} & \floatm{4.21E-06} & \floatm{9.71E-06} & \floatm{1.73E-06} & \floatm{2.08E-05} & \floatm{3.43E-06} & \floatm{9.06E-06} & \floatm{1.99E-06} & \floatm{6.95E-05} & \floatm{1.05E-05} & \floatm{1.85E-05} & \floatm{2.89E-06} & \floatm{2.94E-04} & \floatm{3.23E-05} \\
     & & V100 & \floatm{1.61E-05} & \floatm{1.68E-05} & \floatm{8.25E-06} & \floatm{4.59E-06} & \floatm{1.94E-06} & \floatm{1.51E-06} & \floatm{6.21E-06} & \floatm{3.53E-06} & \floatm{2.22E-06} & \floatm{1.62E-06} & \floatm{2.22E-05} & \floatm{1.13E-05} & \floatm{3.31E-06} & \floatm{2.41E-06} & \floatm{9.12E-05} & \floatm{3.72E-05} \\
     & & A100 & \floatm{3.57E-05} & \floatm{3.39E-05} & \floatm{4.37E-06} & \floatm{3.01E-06} & \floatm{1.56E-06} & \floatm{1.42E-06} & \floatm{4.11E-06} & \floatm{2.44E-06} & \floatm{2.07E-06} & \floatm{1.53E-06} & \floatm{1.25E-05} & \floatm{7.17E-06} & \floatm{2.64E-06} & \floatm{2.25E-06} & \floatm{3.56E-05} & \floatm{2.24E-05} \\
     & & MI250 & \floatm{4.02E-05} & \floatm{3.96E-05} & \floatm{7.74E-06} & \floatm{3.96E-06} & \floatm{1.74E-06} & \floatm{1.41E-06} & \floatm{6.03E-06} & \floatm{3.20E-06} & \floatm{2.00E-06} & \floatm{1.54E-06} & \floatm{3.05E-05} & \floatm{1.88E-05} & \floatm{3.51E-06} & \floatm{2.64E-06} & \floatm{5.50E-05} & \floatm{3.02E-05} \\
     & & VU9P & \dots & \dots & \floatm{3.06E-07} & \dots & \dots & \dots & \dots & \dots & \dots & \dots & \dots & \dots & \dots & \dots & \dots & \dots \\
      \hline
    \multirow{6}{*}{Cu}
     & 
       & EPYC & \floatm{1.14E-06} & \floatm{7.02E-07} & \floatm{2.22E-05} & \floatm{9.38E-06} & \floatm{3.43E-06} & \floatm{2.04E-06} & \floatm{1.19E-05} & \floatm{5.28E-06} & \floatm{3.09E-06} & \floatm{1.56E-06} & \floatm{4.79E-05} & \floatm{1.95E-05} & \floatm{4.20E-06} & \floatm{2.73E-06} & \floatm{2.00E-04} & \floatm{6.21E-05} \\
     & & 3080 Ti & \floatm{1.49E-05} & \floatm{8.98E-06} & \floatm{3.05E-05} & \floatm{4.18E-06} & \floatm{8.52E-06} & \floatm{1.51E-06} & \floatm{1.88E-05} & \floatm{3.15E-06} & \floatm{7.98E-06} & \floatm{1.81E-06} & \floatm{7.46E-05} & \floatm{1.12E-05} & \floatm{1.47E-05} & \floatm{2.32E-06} & \floatm{2.94E-04} & \floatm{3.30E-05} \\
     & & V100 & \floatm{1.57E-05} & \floatm{1.57E-05} & \floatm{8.73E-06} & \floatm{4.81E-06} & \floatm{1.56E-06} & \floatm{1.27E-06} & \floatm{5.71E-06} & \floatm{3.18E-06} & \floatm{1.84E-06} & \floatm{1.38E-06} & \floatm{2.43E-05} & \floatm{1.22E-05} & \floatm{2.60E-06} & \floatm{1.83E-06} & \floatm{9.11E-05} & \floatm{3.73E-05} \\
     & & A100 & \floatm{3.69E-05} & \floatm{3.69E-05} & \floatm{4.41E-06} & \floatm{2.65E-06} & \floatm{1.36E-06} & \floatm{1.15E-06} & \floatm{3.35E-06} & \floatm{2.15E-06} & \floatm{1.63E-06} & \floatm{1.42E-06} & \floatm{1.35E-05} & \floatm{7.49E-06} & \floatm{2.15E-06} & \floatm{1.78E-06} & \floatm{3.62E-05} & \floatm{2.10E-05} \\
     & & MI250 & \floatm{3.90E-05} & \floatm{3.91E-05} & \floatm{8.27E-06} & \floatm{4.13E-06} & \floatm{1.37E-06} & \floatm{1.21E-06} & \floatm{5.62E-06} & \floatm{2.98E-06} & \floatm{1.59E-06} & \floatm{1.35E-06} & \floatm{2.69E-05} & \floatm{1.26E-05} & \floatm{2.56E-06} & \floatm{2.00E-06} & \floatm{5.54E-05} & \floatm{2.95E-05} \\
     & & VU9P & \dots & \dots & \floatm{3.10E-07} & \dots & \dots & \dots & \dots & \dots & \dots & \dots & \dots & \dots & \dots & \dots & \dots & \dots \\
      \hline
    \multirow{5}{*}{HEA}
     & 
       & EPYC & \dots & \dots & \floatm{3.28E-05} & \floatm{1.30E-05} & \floatm{7.04E-06} & \floatm{4.58E-06} & \floatm{1.53E-05} & \floatm{7.64E-06} & \floatm{6.83E-06} & \floatm{3.80E-06} & \floatm{8.10E-05} & \floatm{3.34E-05} & \floatm{8.56E-06} & \floatm{5.68E-06} & \floatm{1.56E-04} & \floatm{4.59E-05} \\
     & & 3080 Ti & \dots & \dots & \floatm{6.53E-05} & \floatm{9.72E-06} & \floatm{1.05E-05} & \floatm{2.51E-06} & \floatm{3.61E-05} & \floatm{6.83E-06} & \floatm{1.19E-05} & \floatm{3.24E-06} & \floatm{1.71E-04} & \floatm{2.49E-05} & \floatm{2.96E-05} & \floatm{5.37E-06} & \floatm{2.90E-04} & \floatm{3.28E-05} \\
     & & V100 & \dots & \dots & \floatm{2.01E-05} & \floatm{1.09E-05} & \floatm{2.88E-06} & \floatm{2.39E-06} & \floatm{1.23E-05} & \floatm{6.86E-06} & \floatm{1.23E-05} & \floatm{2.85E-06} & \floatm{5.52E-05} & \floatm{2.84E-05} & \floatm{9.42E-06} & \floatm{5.47E-06} & \floatm{9.12E-05} & \floatm{3.74E-05} \\
     & & A100 & \dots & \dots & \floatm{1.04E-05} & \floatm{6.09E-06} & \floatm{2.13E-06} & \floatm{1.83E-06} & \floatm{7.25E-06} & \floatm{5.48E-06} & \floatm{2.98E-06} & \floatm{2.83E-06} & \floatm{3.01E-05} & \floatm{1.71E-05} & \floatm{4.21E-06} & \floatm{4.22E-06} & \floatm{3.50E-05} & \floatm{2.00E-05} \\
     & & MI250 & \dots & \dots & \floatm{2.01E-05} & \floatm{1.16E-05} & \floatm{4.57E-06} & \floatm{4.22E-06} & \floatm{1.62E-05} & \floatm{1.20E-05} & \floatm{7.01E-06} & \floatm{6.44E-06} & \floatm{7.60E-05} & \floatm{4.49E-05} & \floatm{9.09E-06} & \floatm{7.61E-06} & \floatm{5.57E-05} & \floatm{3.05E-05} \\
    \hline \hline
    \end{tabular}%

}%
    \justifying
    \noindent
    \label{t:performance-md}\vspace{-0.5cm}
\end{table*}

We present the validation errors of different models in Table~\ref{t:error}, as well as the training and MD performance on various platforms in Table~\ref{t:performance-training} and \ref{t:performance-md}. None of the models outperforms the others in terms of accuracy for all datasets.
The non-smooth local frame descriptor achieves the best accuracy for the water system, with an energy RMSE of 0.689~meV/atom and a force RMSE of 39.2~meV/\AA{}.
Moreover, this model exhibits the fastest computing performance among all models on CPUs, although it has not yet been implemented on GPUs.
The local frame descriptor, despite having higher accuracy in some cases, has limitations that hinder its widespread applicability.
One such limitation is that it is not smooth.
Additionally, this descriptor does not perform well for the copper system, which was collected over a wide range of temperatures and pressures \cite{Zhang_ComputPhysCommun_2020_v253_p107206}. Another limitation is that it requires all systems to have similar chemical species to build the local frame, which makes it challenging to apply in datasets like HEA, OC2M, dipeptides, and SPICE.

On the other hand, the DeepPot-SE descriptor offers greater generalization in terms of both accuracy and performance.
The compressed models are 1x-10x faster than the original for training and inference, and the NVNMD is 50x-100x faster than the regular MD, both of which demonstrate impressive computational performance.
The three-body embedding descriptor theoretically contains more information than the two-body embedding descriptor and is expected to be more accurate but slower.
While this is true for the water and copper systems, the expected order of accuracy is not clearly observed for the HEA and dipeptides datasets.
Further research is required to determine the reason for this discrepancy, but it is likely due to the loss not converging within the same training steps when more chemical species result in more trainable parameters.
Furthermore, the performance on these two datasets slows down as there are more neural networks.

The attention-based models with the type embedding exhibit better accuracy for the HEA system and equivalent accuracy for the dipeptides system.
These models also have the advantage of faster training on GPUs, with equivalent accuracy for these two systems, by reducing the number of neural networks.
However, this advantage is not observed on CPUs or MD simulations, as attention layers are computationally expensive, which calls for future improvements.
Furthermore, when there are many chemical species, the attention-based descriptor requires less CPU or GPU memory than other models since it has fewer neural networks.
This feature makes it possible to apply to the OC2M dataset with over 60 species and the SPICE dataset with about 20 species.

It is noteworthy that in nearly all systems, FP32 is 0.5x to 2x faster than FP64 and demonstrates similar validation errors.
Therefore, FP32 should be widely adopted in most applications.
Moreover, FP32 enables high performance on hardware with poor FP64 performance, such as consumer GPUs or CPUs.

\section{Summary}
DeePMD-kit is a powerful and versatile community-developed open-source software package for molecular dynamics (MD) simulations using machine learning potentials (MLPs).
Its excellent performance, usability, and extensibility have made it a popular choice for researchers in various fields.
DeePMD-kit is licensed under the LGPL-3.0 license, which allows anyone to use, modify, and extend the software freely.
Thanks to its well-designed code architecture, DeePMD-kit is highly customizable and can be easily extended in various aspects. The models are organized as Python modules in an object-oriented design and saved into the computing graphs, making it easier to add new models. The computing graph is composed of TensorFlow and customized operators, making it easier to optimize the package for a particular hardware architecture and certain operators. The package also has rich and flexible APIs, making it easier to integrate with other molecular simulation packages.
DeePMD-kit is open to contributions from researchers in computational science, and we hope that the community will continue to develop and enhance its features in the future.

\section*{Data Availability}

DeePMD-kit is openly hosted at the GitHub repository \url{https://github.com/deepmodeling/deepmd-kit}.
The datasets, the models, the simulation systems, and the benchmarking scripts used in this study can be downloaded from the GitHub repository \url{https://github.com/deepmodeling-activity/deepmd-kit-v2-paper}.
Other data that support the findings of this study are available from the corresponding author upon reasonable request.

\section*{Acknowledgments}

The authors thank  Yihao Liu, Xinzijian Liu, Haidi Wang, Hailin Yang, and the GitHub user ZhengdQin for their code contribution to DeePMD-kit.
D.T. is grateful to Stefano Baroni, Riccardo Bertossa, Federico Grasselli, and Paolo Pegolo for enlightening discussions throughout the completion of this work.
ChatGPT was used to polish the manuscript under supervision.
The work of J.Z. and D.M.Y. is supported by the National Institutes of Health (Grant No.~GM107485 to D.M.Y.) and the National Science Foundation (Grant No.~2209718 to D.M.Y.).
J.Z. is grateful for the Van Dyke Award from the Department of Chemistry and Chemical Biology, Rutgers, The State University of New Jersey.
The work of Y.C., Yifan Li, and R.C. is supported by the ``Chemistry in Solution and at Interfaces'' (CSI) Center funded by the United States Department of Energy Award DE-SC0019394.
The work of M.R. is supported by the VEGA Project No.~1/0640/20 and by the Slovak Research and Development Agency under Contract No.~APVV-19-0371.
The work of Q.Z. is supported by the Science and Technology Innovation Program of Hunan Province under Grant No.~2021RC4026.
The work of S.L.B. was supported by the Research Council of Norway through the Centre of Excellence Hylleraas Centre for Quantum Molecular Sciences (grant number 262695).
The work of C.L. and R.W. is supported by the United States Department of Energy (DOE) Award DE-SC0019759.
The work of H.W. is supported by the National Key R\&D Program of China under Grant No.~2022YFA1004300, and the National Natural Science Foundation of China under Grant No.~12122103.
Computational resources were provided by
the Bohrium Cloud Platform at DP technology;
the Office of Advanced Research Computing (OARC) at Rutgers, The State University of New Jersey;
the Advanced Cyberinfrastructure Coordination Ecosystem: Services \& Support (ACCESS) program, which is supported by National Science Foundation grants \#2138259, \#2138286, \#2138307, \#2137603, and \#2138296 (supercomputer Expanse at SDSC through allocation CHE190067);
the Texas Advanced Computing Center (TACC) at the University of Texas at Austin, URL: http://www.tacc.utexas.edu (supercomputer Frontera through allocation CHE20002);
the AMD Cloud Platform at AMD, Inc;
and the Princeton Research Computing resources at Princeton University, which is a consortium of groups led by the Princeton Institute for Computational Science and Engineering (PICSciE) and the Office of Information Technology’s Research Computing.

\bibliography{reference}

\end{document}